\documentclass[runningheads]{llncs}
\usepackage{amsfonts} 
\usepackage{amsmath,amssymb}
\usepackage{comment}
\usepackage{graphicx}
\usepackage{wrapfig}
\usepackage{subfigure}
\usepackage{tikz}
\usepackage{xcolor}
\usepackage{bm}
\usepackage{mathtools}
\usepackage{fontawesome}
\usepackage{enumitem}
\usepackage{hyperref}

\usepackage{verbatim}
\newcommand{\q}{\textbf{q}}

\newcommand{\daniele}[1]{\textcolor{orange}{[DT: #1]}}

\begin{document} 
\title{Contagion Effects of Heterogeneous Cyber
Risk on Network Security and Systemic Stability 
}

\titlerunning{Abbreviated paper title}

\author{Elisa Botteghi\inst{1}
\and Martino Centonze\inst{1}
\and Davide Pastorello\inst{1,2}
\and Daniele Tantari\inst{1}
}

\institute{Department of Mathematics, University of Bologna \\
\and
TIFPA-INFN, via Sommarive 14, 38123 Povo TN (Italy) 
\email{daniele.tantari@unibo.it} 
}

\maketitle              
\begin{abstract}
Understanding how heterogeneous cyber risk shapes security investments and contagion is a key challenge for the resilience of interconnected digital systems.
Existing studies on cybersecurity investments and network contagion typically rely on homogeneous assets, uniform attack incentives, or non-strategic threat propagation.
We develop a Stackelberg security game that combines contagion dynamics with heterogeneous cyber risk, allowing attackers and defenders to assign different values to network nodes, reflecting differences in
asset criticality, information, and objectives. 
We characterize the equilibrium allocation of cybersecurity investments and derive an analytical approximation for the optimal defense strategy based on endogenous network protection metrics.
We show that cyber-risk heterogeneity significantly affects attack incentives and resource allocation, while neglecting it leads to systematic defense misallocation.
Our results suggest that cybersecurity policies should leverage both actual and perceived differences in node criticality, as cyber-deception mechanisms can create strategic misperceptions that redirect attacks and enhance systemic resilience.

\end{abstract}

\section{Introduction}

Cyber risk refers to the potential for damage to digital systems and infrastructure, whether in terms of functionality or data integrity, resulting from security breaches.

In today’s digital economy, cyber risk is inherently a financial risk for two distinct reasons.  First, cyber-attacks generate direct financial losses.
Operational disruptions and data breaches directly impact assets of economic value \cite{eling2019actual}, as operability itself has financial worth and data are increasingly treated as valuable commodities. Second, cyber risk affects the financial system indirectly. Modern financial systems rely on digital infrastructures \cite{nucci2023digital}, often described as cyber-physical systems. Attacks targeting the digital layer can therefore propagate to the financial layer, transmit risk across institutions, and potentially trigger systemic instability \cite{crosignani2023pirates,jamilov2021anatomy,aldasoro2022drivers}. For example, cyberattacks targeting technology service providers, such as digital payment platforms, may lead to liquidity crises within interbank networks. Additional spillover effects include adverse stock market reactions and reputational damage following cyber incidents \cite{piccotti2023informed,amir2018firms,iyer2020cyberattacks,akey2026hacking}. As a result, effective cyber-risk management and strategic investment in cybersecurity are essential not only for digital resilience but also for the broader stability of financial systems.

As digital ecosystems become increasingly complex, the need to manage large amounts of data has led to infrastructures that are typically organized as networks of interconnected servers or units. While such interconnectivity enhances efficiency, scalability, and redundancy, it also introduces systemic vulnerabilities: a single breach can act as a gateway for contagion, allowing malicious activity to propagate and potentially compromise the entire system.

A key insight from the economics of information security \cite{anderson2006economics} is that heterogeneity is a fundamental feature of real-world systems and a primary driver of both vulnerability and resilience. Digital infrastructures are characterized by highly uneven patterns of connectivity, asymmetric valuations of assets, and attackers who selectively target components that yield the highest expected payoff. In such environments, security failures rarely result from random shocks: instead, they emerge from strategic behavior that exploits heterogeneity in system structure and node importance. Some nodes are disproportionately critical, either because of their connectivity, their economic value, or their role within the system, while others can be compromised with limited systemic impact. Similarly, attacks differ substantially in scope and objective, targeting specific services, data, or functionalities rather than spreading uniformly. These observations suggest that models assuming homogeneous networks, identical nodes, or indiscriminate attacks risk overlooking key mechanisms shaping systemic cyber risk and may lead to misleading prescriptions for optimal defense. These considerations motivate the need for a modeling framework that explicitly incorporates heterogeneity at multiple levels.

In this work, we propose a cyber risk management framework for heterogeneous digital networked systems, designed to identify an optimal investment strategy for allocating cybersecurity defenses across network nodes. The core components of our model include contagion across the network and the strategic behavior of cyber attacks.

For the first component, we draw inspiration from the extensive literature on complex networks \cite{newman2018networks}. Complex network models of contagion have been widely applied to biological \cite{pastor2015epidemic}, social \cite{nekovee2007theory}, and financial systems \cite{gai2010contagion,acemoglu2015systemic,agosto2020tree,macdonald2015investigation,chen2024contagion,ATASOY2024106596,calice2023contingent}, where the propagation of shocks or behaviors is mediated by the structure of interactions among agents.
Typically, these models assume that the initial trigger, be it the zero-patient of an epidemic, a financial shock or a rumor, is exogenous and random, reflecting events that are unintentional and difficult to predict or control.

However, this assumption does not hold in the context of cybersecurity, where attacks are often strategic and targeted. To address this, we adopt a game-theoretic framework, building on the literature of security games \cite{tambe2011security,liang2012game}, which explicitly models the adversarial nature of cyber threats and the strategic allocation of defenses.

In particular, two players Stackelberg security games \cite{sinha2018stackelberg}, in which a defender (leader) commits to a defensive strategy and an attacker (follower) responds accordingly,  have received significant interest due to their wide range of applications including patrolling tasks \cite{pita2008deployed}, environmental protection efforts \cite{fang2015security,xepapadeas2023multi} and defense coordination \cite{gan2022defense}. Within this framework, security games that incorporate interdependencies among nodes, where insufficient protection of one node can undermine the defense of its neighbors, have been studied in \cite{LouSmithVorobeychik2017,ChanCeykoOrtiz2017,LiTranThanhWu2020}. 

Models that explicitly account for contagion phenomena, where threats propagate across the network, have been also examined in \cite{aspnes2006inoculation,acemoglu2016network,tsai2012security,bai2023stackelberg,GoyalVigier2014}. Most of these models, still grounded in an epidemiological view of contagion, assume that the attacker's objective is simply to maximize the spread of infection across the network. However, in cybersecurity settings, the attacker may instead aim to reach specific nodes or regions considered strategic or of higher value. Similarly, the defender may prioritize the protection of critical nodes while allowing others to be sacrificed.

In our work, building on the contagion mechanism of \cite{aspnes2006inoculation,acemoglu2016network}, we therefore consider a heterogeneous network in which both attacker and defender assign values to each node and a corresponding target risk. The attacker and defender may assign different value/risk profiles to the nodes, reflecting their asymmetric information and differing perspectives on the system's structure and strategic importance. The two players compete within a security game with limited budget: the defender strategically allocates cybersecurity resources across the network nodes, while the attacker determines an optimal probability distribution over potential targets.

The economic mechanism underlying our framework is that cyber-risk heterogeneity shapes both attack incentives and defensive resource allocation. When attackers and defenders assign different values to network assets, cybersecurity investments are no longer determined solely by the topological importance of nodes. Instead, optimal protection emerges from the interaction between network structure and heterogeneous perceptions of node criticality. As a result, resources may be strategically concentrated on nodes that are not necessarily the most connected, but that provide the greatest contribution to overall system resilience against targeted attacks. This highlights that heterogeneity is not merely a descriptive feature of cyber systems, but a key determinant of efficient cybersecurity investment decisions.

The main contributions of this paper are organized as follows. In Section \ref{sec:model}, we introduce the general setting. Section \ref{sec:NetworkContagion} defines the static contagion mechanism adopted. In Section \ref{sec:model_Game}, we present the security game framework, including the attacker’s and defender’s value/risk profiles. Section \ref{sec:mod_stack} recalls the Stackelberg formulation of the game and the corresponding definition of  Strong Stackelberg Equilibria.

Section \ref{sec:result} outlines the main analytical results. In Section \ref{sec:thm}, we provide a characterization of the Stackelberg equilibrium based on intuitive and naturally emerging network metrics. These metrics quantify the extent to which each node contributes to the overall protection of the system and are combined in a principled way according to the players’ respective risk profiles. Section \ref{sec:path} introduces a class of risk measures defined in terms of the number of contagion paths, allowing for the inclusion of a dynamic parameter linked to propagation time.

Section \ref{sec:num} discusses the numerical results. In Section \ref{sec:effront}, we use the efficient frontier to compare the overall system robustness to strategic attacks across different network topologies and risk profiles. Section \ref{sec:dec} analyzes examples of optimal resource allocation patterns, with a focus on the emergence of cyber deception effects. In Section \ref{sec:dyn}, we examine the robustness of the optimal strategies to model misspecifications across a range of different dynamic contagion mechanisms.

Finally, Section \ref{sec:conclusions} presents the conclusions and outlines future research directions. The proof of the main theorem is provided in Appendix \ref{sec:app}.

\subsection{Background and related literature}

Our contribution relates to the literature on the economics of cybersecurity, with a particular focus on cybersecurity investment decisions among interconnected firms \cite{fedele2022dangerous}.

Following the seminal work of Gordon and Loeb \cite{gordon2002economics}, early economic models of cybersecurity focused on firms’ investment incentives using single-firm frameworks, thus neglecting interdependencies among firms. Yet, firms typically operate on shared computer networks, leading to technical spillovers \cite{kunreuther2003interdependent,varian2004system}, and often compete in product markets, giving rise to market spillovers \cite{garcia2007potential,liao2014network,arce2018malware}. As a result, the literature has naturally evolved toward the analysis of security in interconnected systems.

The primary objective of this literature is to compare firms’ equilibrium cybersecurity investments in decentralized settings with the socially optimal investment level achievable under centralized coordination. In networked environments, individual investments create positive externalities by improving the security of interconnected firms. Yet, these externalities can induce free-riding incentives, whereby firms rely on others’ investments rather than investing themselves \cite{varian2004system}, resulting in underinvestment relative to the social optimum \cite{kunreuther2003interdependent}.

Dziubiński and Goyal \cite{dziubinski2017you} show that when cybersecurity investments are strategically complementary, firms may invest only if enough other firms do so. This feature leads to equilibria with no investment and can generate coordination failures.



    Building on earlier contributions, Acemoglu et al. \cite{acemoglu2016network} model cybersecurity decisions in the presence of a strategic attacker who conditions attacks on firms’ investment levels rather than acting randomly. This feature creates incentives for firms to overinvest in security to divert attacks toward other firms, generating a negative externality and yielding overinvestment as a key implication.

All the aforementioned contributions typically assume that the network topology is exogenously given. By contrast, a parallel strand of the literature investigates how network structures endogenously emerge as the outcome of security games. Dziubiński and Goyal \cite{dziubinski2013network} study a two-stage game in which a centralized defender first designs the network and allocates defensive resources, and an attacker subsequently chooses which firms to target. They show that equilibrium network density depends on the cost of defense. Allowing for contagion, Goyal and Vigier \cite{GoyalVigier2014} show that equilibrium networks may feature either a single central hub or multiple hubs, depending on model parameters. Cerdeiro et al. \cite{cerdeiro2017individual} extends the framework to decentralized defense and analyze how network topology can mitigate the distorted effect of positive, negative and coordination externalities on firms’ incentives to invest in cybersecurity.

Garcia et al. \cite{garcia2007potential} is among the first to analyze interdependence arising from market spillovers. In his framework, a firm hit by a cyberattack loses market share, which is instead captured by competitors that remain unaffected. Liu et al. \cite{liu2018security} consider the opposite situation of two firms selling complementary products and show that a cyberattack on one firm reduces demand for the other, reflecting the absence of substitutability between the goods.

Our framework extends the one considered by Acemoglu et al. \cite{acemoglu2016network} and shares several similarities with Aspnes et al. \cite{aspnes2006inoculation}. The latter study antivirus immunization in a network subject to contagious random attacks under decentralized defense. They characterize Nash equilibria in terms of a critical threshold in the size of each node’s connected component.

Focusing on centralized defense, our contribution is also closely related to the computer science literature on security in computer and communication networks, which analyzes the protection of digital infrastructures composed of interconnected nodes. Within this interpretation, the central authority corresponds to a single firm, the network represents its internal digital infrastructure, and the defense strategy reflects the optimal allocation of cybersecurity resources across internal units.

In contrast to our framework, Li et al. \cite{LiTranThanhWu2020} considers a setting in which defensive resources allocated to a node can be partially shared with its neighbors, with each edge weighted to capture the efficiency of resource sharing between adjacent nodes. By contrast, Bai et al. \cite{bai2023stackelberg} argue that sharing defensive resources inherently reduces the defensive strength of the source node. Accordingly, they model a reallocation problem in which resources can be transferred to neighboring nodes at a cost. Finally, Lou et al. \cite{LouSmithVorobeychik2017} study node interdependencies in a multi-defender security game, where each defender can simultaneously protect multiple nodes. Multi-step attacks propagating through the system have been extensively studied using the formalism of attack graphs \cite{khouzani2019scalable}.

All the contributions discussed above rely on some degree of homogeneity in the system. Specifically, they either assume homogeneous interdependencies among firms or posit that system units are identical in terms of value and potential losses when exposed to cyber threats, thus contributing equally to social welfare. Moreover, attacks are typically modeled as either uniformly distributed or as the outcome of strategic interactions in which all units are equally valuable targets for the attacker.

These assumptions of homogeneity are difficult to reconcile with real-world systems: firms may be connected through highly heterogeneous network structures and interdependencies, as pointed in \cite{anderson2006economics}. Furthermore,  from the perspective of a central authority, different part of the system often exhibit markedly different economic values and acceptable risk levels. In addition, cyber threats are inherently heterogeneous, ranging from denial-of-service and phishing attacks to ransomware and password breaches, each targeting distinct components of the system. The proposed framework addresses this gap by introducing a security game between a central authority managing a system of interconnected units and a strategic attacker, in which both the network topology and the valuation of individual units are heterogeneous from the perspectives of both the defender and the attacker. Our main contribution is the derivation of an efficient expression for the optimal cybersecurity investment strategy, which jointly captures and highlights the distinct roles of these heterogeneities and their interaction.

While we focus on security investments and their optimal allocation, other relevant approaches in cybersecurity focus on temporal
adaptation of the system to the attack. 

Moving Target Defense (MTD) refers to cybersecurity techniques that dynamically modify system configurations to increase uncertainty and reduce the attacker’s probability of success. MTD ha been extensively studied thoruogh game-theoretic models, where defender strategically reconfigures the system over time. In this context, the attacker-defender interaction is often formulated as a stochastic game capturing the sequential nature of cyber-conflicts \cite{TAN2023100544}.

From a different perspective, the literature on PIoT PLC honeypot mechanisms studies defensive strategies based on deception. In this setting, defenders deploy simulated programmable logic controllers (PLCs) to attract and mislead attackers, allowing them to detect, divert, or analyze malicious behavior without exposing real assets to risk. These approaches are typically modeled using time-delay evolutionary game frameworks \cite{11217210}.

MTD and PLC honeypoint techniques share common features with our framework, including the use of game-theoretic models and the role of deception in selecting optimal defense strategies. However, they often abstract away from the role of the network topology and overlook the economic dimensions of the problem.

\section{Model}\label{sec:model}

\subsection{Contagion on Networks}\label{sec:NetworkContagion}

Let us consider an undirected network $\mathcal{G}$ of $n$ interconnected nodes represented by its adjacency matrix $\mathbf{A}=(A_{ij})_{i<j}^n$, $A_{ij}\in \{0,1\}$, $i,j=1,\ldots,n$. Each entry $A_{ij}$ encodes the presence ($A_{ij}=1$) or absence ($A_{ij}=0$) of a link between the pair of nodes $(i,j)$.

Each node $s$ can be the target of a cyber-threat with a probability $\phi_s\in[0,1]$,  becoming the \textit{seed} of a potential infection. Therefore, the cyber-attack distribution is represented by the vector $\bm{\phi}=(\phi_i)_{i=1}^n$, $\sum_{i=1}^n \phi_i=1$.

Nodes can be either immune or susceptible to cyber-attacks. We introduce the susceptibility vector $\bm{X} = (X_i)_{i=1}^n$, where each $X_i \in \{0,1\}$ indicates whether node $i$ is susceptible ($X_i = 1$) or immune ($X_i = 0$) to a cyber-threat. We model $\bm{X}$ as a random vector of independent Bernoulli variables, with $\mathbb{P}(X_i = 0) = q_i$ and $\mathbb{P}(X_i = 1) = 1 - q_i$, where $q_i \in [0,1]$. The vector $\bm{q} = (q_i)_{i=1}^n$ represents the \textit{security level} of the system, with $q_i$ denoting the security level of node $i$. The vector $\bm{q}$ reflects how cybersecurity investments have been distributed across the network. 

All model stochasticity is encoded in the probability space defined by the set of elementary events $ (s,\bm{X})\in \mathbb{N}_n\times \{0,1\}^n =:\Omega$ together with the joint probability distribution
\begin{equation}\label{def:p(s,X)}
p(s,\bm{X}):= \phi_s \prod_{i=1}^n (1-q_i)^{X_i}q_i^{1-X_i},
\end{equation}
where we assume that the seed location and the susceptibilities are independent random variables.

A security breach at a node poses a threat to its neighbors, potentially allowing a cyber-infection to spread throughout the entire system, starting from the initial seed. Since the infection propagates only through susceptible nodes, any susceptible node that is connected to the seed via a path consisting entirely of susceptible nodes will become infected. Alternatively, we can define the \textit{transmission network} $\mathcal{T}(\bm{X})$ as the (random) sub-network of $\mathcal{G}$ consisting only of susceptible nodes. Then, a node $i$ becomes infected if and only if it belongs to the same connected component as the seed $s$ in the transmission network, i.e. if there exists a path $i \sim s$ between $i$ and $s$ entirely in $\mathcal{T}(\bm{X})$.  The infection condition can be simply expressed by the following
\begin{definition}[Infected nodes]\label{def:infection}
 Given a realization of the random vector $\bm{X}$ and a seed $s$, a node $i$ is said to become infected if there exists $\ell \in \mathbb{N}$ such that
\begin{equation}
\left( \bm{A} \circ \bm{X}\bm{X}^T \right)^{\ell}_{is} \geq 1,\nonumber
\end{equation}
where $\circ$ denotes the element-wise (Hadamard) product, i.e., $\left( \bm{A} \circ \bm{X}\bm{X}^T \right)_{ij} = A_{ij} X_i X_j$.
\end{definition}
Let us stress that $\bm A\circ \bm X\bm X^T$ is the adjacency matrix of $\mathcal T(\bm X)$ so $\left( \bm{A} \circ \bm{X}\bm{X}^T \right)^{\ell}_{is}$ is the number of paths of length $l$ from $s$ to $i$ in $\mathcal T(\bm X)$.

\begin{figure}
\centering
\begin{tikzpicture}[scale=1.5, every node/.style={circle, draw, minimum size=3mm}]

\node[fill=orange!50] (A) at (0,0) {};

\node[fill=blue!20] (B) at (0.1,-1) {};
\node[fill=orange!50] (C) at (0.05,-2) {};
\node[fill=blue!20] (D) at (-0.5,-2.3) {};
\node[fill=blue!20] (E) at (-1,-2.6) {};
\node[fill=blue!20] (F) at (0.1,-2.3) {};
\node[fill=blue!20] (F') at (-0.1,-2.6) {};

\node[fill=orange!50] (G) at (-0.25, 0.75) {};
\node[fill=blue!20] (H) at (-0.75, 1) {};
\node[fill=orange!50] (I) at (-1, 1.25) {};
\node[fill=orange!50] (J) at (-1.25, 1) {};
\node[fill=blue!20] (J') at (-1.5, 1.25) {};
\node[fill=blue!20] (I') at (-1.25, 1.5) {};

\node[fill=orange!50] (K) at (0.25, 0.75) {};
\node[fill=orange!50]   (L) at (0.75, 1.2) {};
\node[fill=blue!20]   (M) at (0.75, 0.75) {};
\node[fill=blue!20]   (M1) at (1.2, 1) {};
\node[fill=orange!50]   (M2) at (1.2, 1.5) {};
\node[fill=blue!20]   (M3) at (0.75, 1.7) {};
\node[fill=blue!20]   (M4) at (0.35, 1.3) {};

\node[fill=orange!50] (N) at (-0.45, -0.1) {};
\node[fill=blue!20]   (O) at (-0.75, -0.65) {};
\node[fill=orange!50]   (O1) at (-1, 0.25) {};
\node[fill=orange!50]   (O2) at (-1.25, -0.2) {};

\draw[-] (A)--(B);
\draw[-] (B)--(C);
\draw[-] (C)--(D);
\draw[-] (D)--(E);
\draw[-] (C)--(F);
\draw[-] (F)--(F');

\draw[-] (A)--(G);
\draw[-] (G)--(H);
\draw[-] (H)--(I);
\draw[-] (I)--(J);
\draw[-] (J')--(J);
\draw[-] (J')--(I');
\draw[-] (J)--(I');
\draw[-] (I)--(I');

\draw[-] (G)--(K);
\draw[-] (K)--(L);
\draw[-] (L)--(M);
\draw[-] (L)--(M1);
\draw[-] (L)--(M2);
\draw[-] (L)--(M3);
\draw[-] (L)--(M4);

\draw[-] (A)--(N);
\draw[-] (N)--(O);
\draw[-] (O)--(O1);
\draw[-] (O1)--(O2);
\draw[-] (O)--(O2);

\begin{scope}[xshift=3.5cm]

\node[fill=red!50] (A) at (0,0) {};

\node[fill=blue!20,inner sep=0pt] (B) at (0.1,-1) {$\bm{\times}$};
\node[fill=orange!50] (C) at (0.05,-2) {};
\node[fill=blue!20] (D) at (-0.5,-2.3) {};
\node[fill=blue!20] (E) at (-1,-2.6) {};
\node[fill=blue!20] (F) at (0.1,-2.3) {};
\node[fill=blue!20] (F') at (-0.1,-2.6) {};

\node[fill=red!50] (G) at (-0.25, 0.75) {};
\node[fill=blue!20,inner sep=0pt] (H) at (-0.75, 1) {$\bm{\times}$};
\node[fill=orange!50] (I) at (-1, 1.25) {};
\node[fill=orange!50] (J) at (-1.25, 1) {};
\node[fill=blue!20] (J') at (-1.5, 1.25) {};
\node[fill=blue!20] (I') at (-1.25, 1.5) {};

\node[fill=red!50] (K) at (0.25, 0.75) {};
\node[fill=red!50]   (L) at (0.75, 1.2) {};
\node[fill=blue!20]   (M) at (0.75, 0.75) {};
\node[fill=blue!20]   (M1) at (1.2, 1) {};
\node[fill=red!50]   (M2) at (1.2, 1.5) {\faBolt};
\node[fill=blue!20]   (M3) at (0.75, 1.7) {};
\node[fill=blue!20]   (M4) at (0.35, 1.3) {};

\node[fill=red!50] (N) at (-0.45, -0.1) {};
\node[fill=blue!20,inner sep=0pt]   (O) at (-0.75, -0.65) {$\bm{\times}$};
\node[fill=orange!50]   (O1) at (-1, 0.25) {};
\node[fill=orange!50]   (O2) at (-1.25, -0.2) {};

\draw[-] (A)--(B);
\draw[-] (B)--(C);
\draw[-] (C)--(D);
\draw[-] (D)--(E);
\draw[-] (C)--(F);
\draw[-] (F)--(F');

\draw[-] (A)--(G);
\draw[-] (G)--(H);
\draw[-] (H)--(I);
\draw[-] (I)--(J);
\draw[-] (J')--(J);
\draw[-] (J')--(I');
\draw[-] (J)--(I');
\draw[-] (I)--(I');

\draw[-] (G)--(K);
\draw[-] (K)--(L);
\draw[-] (L)--(M);
\draw[-] (L)--(M1);
\draw[-] (L)--(M2);
\draw[-] (L)--(M3);
\draw[-] (L)--(M4);

\draw[-] (A)--(N);
\draw[-] (N)--(O);
\draw[-] (O)--(O1);
\draw[-] (O1)--(O2);
\draw[-] (O)--(O2);
\end{scope}

\end{tikzpicture}
\caption{Representation of the contagion process. Left panel: a realization of the network showing susceptible nodes (orange) and immune nodes (blue) before the attack; Right panel: after the attack all the susceptible nodes connected to the seed become infected (red). Strategically placed immune nodes can block the spread and protect other susceptible nodes from infection. \label{fig:suscinf}}
\end{figure}

\subsection{Security Game Framework}\label{sec:model_Game}

On top of this environment we consider a two players extensive game involving an attacker and a defender. The attacker optimizes the cyber-attack distribution vector $\bm{\phi}$, strategically selecting the nodes where it is most advantageous to pose a threat. The defender allocates cyber defenses across the network by adjusting the system’s security vector $\bm{q}$, aiming to identify an optimal cybersecurity investment strategy. Therefore, we refer to $\bm q$ and $\bm\phi$ as defender and attacker {\em strategies} respectively.

The attacker’s utility function $\mathcal{U}_a$ and the defender’s loss function $\mathcal{L}_d$, which are to be optimized, are defined with the following risk/cost structure:
\begin{align}
     \mathcal U_{a}(\bm{\phi};\bm{q})&=\sum_{i}^n \eta_i \mathcal{R}_i(\vec q,\vec \phi;\bm{A}) - \theta\  \mathcal C_a(\vec \phi),\label{eq:utility} \\
     \mathcal L_{d}(\vec q;\vec \phi)&=\sum_{i}^n z_i \mathcal{R}_i(\vec q,\vec \phi;\bm{A}) + \alpha \ \mathcal C_d(\vec q). \label{eq:loss}
\end{align}
The strategy costs incurred by the attacker and the defender are denoted by $\mathcal{C}_a(\bm{\phi})$ and $\mathcal{C}_d(\bm{q})$, respectively. While $\mathcal{C}_d(\bm{q})$ concretely measures the size of investments in cyber defenses, $\mathcal{C}_a(\bm{\phi})$ has to be interpreted more as an information cost that penalizes low entropy attacking strategies concentrated on a limited subset of nodes. In the following, we assume either quadratic cost functions of the form $\mathcal{C}_d(\bm{q}) = \frac1 2 \sum_{i=1}^n q_i^2$ and $\mathcal{C}_a(\bm{\phi}) =\frac 1 2\sum_{i=1}^n \phi_i^2$, for analytical tractability, or alternatively $L^1$-type penalizations to promote strategy sparsity. The hyperparameters $\alpha$ and $\theta$ control the relative weight assigned to the cost terms and can be interpreted as Lagrange multipliers in a constrained optimization framework, where the attacker and the defender operate under limited budget constraints.

The vector $\bm{\mathcal{R}}=(\mathcal{R}_i)_{i=1}^n$ in $(\ref{eq:utility}-\ref{eq:loss})$ measures the infection risk at each node, as a function of both the network structure $\bm{A}$ and the strategies adopted by the two players $(\bm{q},\bm{\phi})$, thereby defining  the system's \textit{risk profile}. In the following, we analyze two distinct risk measures: the first is based on the actual probability of a node being infected, while the second relies on the expected number of paths connecting the node to the infection seed. The former enhances the interpretability of the optimal solution, whereas the latter improves its scalability.

Most models addressing contagion dynamics on complex networks are inspired by epidemiological frameworks. They often tacitly assume that the attacker’s objective, either explicitly or as a natural consequence of viral replication, is to maximize damage to the network, merely intended as the overall size of the infection. In a cybersecurity context, a networked system is a heterogeneous collection of assets with varying levels of value. Consequently, each node contributes differently to the overall damage assessment. Moreover, the attacker and the defender may assign different values to the same node, either due to conflicting preferences or because of an information asymmetry between the node’s actual value and the attacker’s perceived value. For this reason, we introduced in $(\ref{eq:utility}-\ref{eq:loss})$ the \textit{value profiles} $\bm{\eta}=(\eta_i)_{i=1}^n, \bm{z}=(z_i)_{i=1}^n \in \mathbb{R}^n$ to model the perceived importance of different nodes or regions of the network for the attacker and the defender, respectively. For example one may consider the case of uniform value profiles
$\bm\eta=\bm z=\bm{1}$,
up to the extreme scenario where the profiles are maximally concentrated on single highly strategic nodes, i.e.
$\bm\eta= \bm{e}_i$, $\bm z= \bm{e}_k$ for some $i,k=1,\ldots,n$.   As before, $\bm{\eta}$ and $\bm{z}$ can be interpreted as Lagrange multipliers in a constrained optimization problem in which the two players operate targeting a given heterogeneous risk profile.

In practical applications, the defender value profile $\bm{z}$ can be constructed from 
empirical cyber-risk indicators associated with each node of the system. 
Typical inputs include 
(i) business--impact assessments such as service-criticality ratings or asset-value scores 
used in Business Impact Analysis (BIA) \cite{torabi2014new}, 
(ii) historical incident frequencies, capturing how often a component has been affected by 
cyber events \cite{shevchenko2023nature}, and 
(iii) operational risk metrics such as financial exposure, loss--given--incident estimates, 
or compliance--driven priority classes \cite{chorafas2003operational}. 
Given a heterogeneous set of such indicators, a composite score can be defined by weighting the relative importance assigned 
to functional criticality, historical exposure, and economic impact. 
\begin{example}
To illustrate the interpretability and practical applicability of the proposed framework, 
we present a small-scale case study on a 10-node enterprise network. We consider a network representative of a small corporate infrastructure, with nodes described in Fig. \ref{fig:casenet}.

\begin{figure}[t]
\centering
\begin{tikzpicture}[scale=0.7]
\tikzset{
  core/.style   ={circle, draw, fill=blue!20,  minimum size=3.5mm, inner sep=0pt},
  entry/.style  ={circle, draw, fill=blue!20,  minimum size=3.5mm, inner sep=0pt},
  crit/.style   ={circle, draw, fill=blue!20,     minimum size=3.5mm, inner sep=0pt},
  mon/.style    ={circle, draw, fill=blue!20,    minimum size=3.5mm, inner sep=0pt},
  edge/.style   ={thick}
}


\node[entry, label={[label distance=1mm]above:1}] (n1) at (-3,  2) {};
\node[core,  label={[label distance=1mm]left:4}] (n4) at (-3,  0) {};

\node[core,  label={[label distance=1mm]above:3}] (n3) at ( 0,  0)  {};
\node[crit,  label={[label distance=1mm]right:9}] (n9) at ( 2.2, 0) {};

\node[core,  label={[label distance=1mm]left:5}] (n5) at (-3, -2) {};
\node[core,  label={[label distance=1mm]below:6}] (n6) at ( 0, -2) {};
\node[crit,  label={[label distance=1mm]right:7}] (n7) at ( 2.2,-2) {};

\node[entry, label={[label distance=1mm]left:2}]  (n2) at (-3, -4) {};
\node[mon,   label={[label distance=1mm]below:10}] (n10) at (-1, -4) {};

\draw[edge] (n4)--(n3);
\draw[edge] (n3)--(n6);
\draw[edge] (n6)--(n5);
\draw[edge] (n5)--(n4);

\draw[edge] (n1)--(n4);
\draw[edge] (n2)--(n5);

\draw[edge] (n3)--(n9);
\draw[edge] (n6)--(n7);

\draw[edge] (n10)--(n5);
\draw[edge] (n10)--(n6);

\end{tikzpicture}

\bigskip

\begin{tabular}{l c c c c}
\hline
\textbf{Node} & \textbf{Asset Type} & \textbf{Criticality} &
\textbf{Incident Freq.}  &
\textbf{$z$} \\
\hline
1  & Web Gateway                 & 3 & 4  & 0.092 \\
2  & Email Server                & 3 & 7  & 0.160 \\
3  & Customer Database           & 5 & 3  & 0.115 \\
4  & Application Server          & 4 & 5  & 0.153 \\
5  & Active Directory / Auth     & 5 & 2  & 0.076 \\
6  & File Server                 & 4 & 2  & 0.061 \\
7  & Backup Server               & 5 & 1  & 0.038 \\
8  & Developer Workstation       & 2 & 9  & 0.137 \\
9  & Finance Workstation         & 4 & 4  & 0.122 \\
10 & Monitoring / Logging Server & 3 & 2  & 0.046 \\
\hline
\end{tabular}
\caption{Schematic 10-node topology: core (3–4–5–6), entry points (1,2), critical periphery (7,9), and monitoring (10). The table shows the associated defender value profile $\bm{z}$ constructed from BIA indicators and historical incident frequencies. \label{fig:casenet}}
\end{figure}

To construct a plausible network topology, we adopt a structure that resembles typical organizational IT architectures. The internal core is formed by nodes 3--4--5--6, which create a central square of tightly interconnected services representing key operational components. Two externally exposed nodes, 1 and 2, serve as natural entry points into the system: node 1 connects to node 4, while node 2 connects to node 5, mirroring the common role of web and email gateways as initial exposure surfaces. Beyond the core, the topology includes several peripheral nodes with specialized but strategically important functions. Node 7 is connected to node 6 and represents the backup infrastructure, which holds high long-term value despite its peripheral position. Node 9, linked to node 3, corresponds to a finance workstation or subsystem that depends directly on the central database. Finally, node 10 acts as a monitoring and logging system, connected to both nodes 5 and 6 to reflect its visibility over authentication flows and system-level activities. To construct the defender valuation vector $z$, we combine two operational quantities:
business-criticality ratings and incident frequencies.  
We assign each node a raw score $
\text{Score}_i = \text{Criticality}_i\  \times \ \text{IncidentFreq}_i$, and obtain $\bm{z}$ by normalizing these values to sum to one.
\end{example}

In many realistic cybersecurity settings, the attacker does not possess perfect information about the network, its architecture, or the true economic value of its components. To capture this, the attacker’s value profile $\bm{\eta}$ should be interpreted as a potentially biased or incomplete perception of node importance. Such misperception may arise from incomplete reconnaissance, where the attacker lacks full visibility of dependencies or systemic roles; from misinformation or deception, for example through the presence of honeypots, decoy nodes, or intentionally obfuscated topologies; or from cognitive and resource constraints, which force the attacker to rely on coarse-grained heuristics rather than precise valuations. As a result, discrepancies between  $\bm{z}$ and $\bm{\eta}$ naturally encode heterogeneous threat scenarios in which perceived and actual criticality differ. Examples of discrepancy classes include:
\begin{itemize}[label=\textbullet]
\item  \emph{Attacker-focused hotspots}, where $\bm{\eta}$ is highly concentrated while $\bm{z}$ is diffuse, reflecting targeted campaigns such as spear-phishing or data-exfiltration operations in which the attacker focuses intensely on a subset of nodes that are not necessarily the most important for the defender.
\item  \emph{Defender-focused hotspots}, where $\bm{z}$ is concentrated and $\bm{\eta}$ is diffuse, arising when certain nodes hold high systemic or economic relevance but the attacker does not initially recognize their importance.
\item  \emph{Strong asymmetry}, where $\bm{\eta}$ and $\bm{z}$ are concentrated on disjoint regions of the network, capturing scenarios dominated by cyber deception, decoy networks, or systematic overvaluation of non-critical assets by the attacker.
\end{itemize}
Different shapes of  $\bm{\eta}$ may also reflect distinct categories of adversaries as characterized in TAL (Threat Agent Library) profiles, such as data miners, insider spies, cyber‑criminal thieves, or politically motivated terrorists, each operating with different objectives, capabilities, and levels of information about the network.

\subsection{Stackelberg Equilibrium}\label{sec:mod_stack}

An extensive game with perfect information is a model for sequential decision-making in which players act in a prescribed order and have complete knowledge of all previous actions taken by others. In particular, we consider a Stackelberg extensive game, where the defender moves first (\textit{leader}) and the attacker makes their decision accordingly (\textit{follower}).

In this setup, while the defender' s strategy is defined by the choice of an action $\q\in S_d= [0,1]^n$, the attacker's strategy is determined by a response function 
\begin{equation}
    \phi: S_d\ni \q \rightarrow \phi(\q)\in S_a,
\end{equation}
that assigns an action to any possible choice of the defender, where $S_a$ is the simplex of probability distributions over $n$ nodes representing all the possible actions of the attacker. We denote the set of possible response functions as $R_a$.

The proposed ordering is natural in our interpretation of the defender's action as a cyber-security investment aimed at strengthening system resilience ahead of potential attacks. Conversely, the opposite ordering would be more appropriate to model scenarios in which the defender, reacting to an attack that has already occurred, actively executes a defensive strategy to block it or mitigate its impact on the system. Although not addressed in the present paper, there are several situations in which it is worthwhile to consider a Stackelberg formulation where the attacker moves first. For example in the case of a {\em ransomwere attack} through the network, the defender reacts by isolating compromised nodes or restoring data from backup \cite{g14020020}. In computer networks, after a {\em data exfiltration attack}, the defender reacts blocking nodes or filtering the suspicious traffic acting on links \cite{10.1007/978-3-319-68711-7_9}. In the context of {dynamic resource allocation}, after a {\em distributed denial-of-service} moved saturating the network resources, the defender reacts reallocating resources \cite{s18114074}. Even {\em cyber kill chain} can be an example of Stackelberg attacker-first framework: the attacker starts with the first move sending malicious messages through network channels then the defender responds with countermeasures over the network such as blocking suspicious domains \cite{kour2025modelling}. 

In a Stackelberg extensive game, a natural definition for an optimal pair of strategies $(\bm{q}^*,\bm{\phi}^*(\cdot))$ is given by the following

 \begin{definition}[Strong Stackelberg equilibrium  SSE]\label{def:SSE}
    In a Stackelberg game with attacker utility functions $\mathcal{U}_a$ and defender loss function $\mathcal{L}_d$, a pair of strategies $(\q^*,\bm{\phi}^*(\cdot))$, where $\q^*\in S_d$ is the defender's strategy and $R_a\ni\bm{\phi}^*:S_d\rightarrow S_a$ is the attacker's response function, is a strong Stackelberg equilibrium  (SSE) if:
     \begin{enumerate}
         \item The attacker plays a best response in each subgame: $$\mathcal{U}_a(\bm{\phi}^*(\q);\q)\geq \mathcal{U}_a(\bm{\phi}(\q);\q), \ \ \ \forall \q\in S_d, \forall \bm{\phi}\in R_a;$$ 
         \item The defender plays a best response: 
         $$\mathcal{L}_d(\q^*;\bm{\phi}^*(\q^*))\leq \mathcal{L}_d(\q;\bm{\phi}^*(\q)),\ \ \ \ \forall \q\in S_d;\ \ \ \ \ \ \ \ \ \ \ \ $$
         \item The attacker breaks ties optimality for the defender: $$\mathcal{L}_d(\q,\bm{\phi}^*(\q))\leq \mathcal{L}_d(\q,\bm{f}^*), \ \ \ \ \ \forall \q\in S_d, \bm{f}^*\in S^\q_{a},$$
     \end{enumerate}
     where we denoted  $S^\q_{a}$ as the set of the attacker best responses to $\q$, i.e. 
     $S^\q_{a}=\{\bm{f}^*\in S_a\, :\,\bm{f}^*=\text{\em arg}\max_{\bm{\phi}\in S_a}\mathcal{U}_a(\bm{\phi};\q)\}.$
 \end{definition}
By definition, a Strong Stackelberg Equilibrium (SSE) is always a subgame perfect equilibrium \cite{ferguson2020course} due to conditions 1 and 2. In the presence of multiple equilibria, given by the existence of multiple best response functions, condition 3 serves to select a subset of them. In general the Nash equilibrium of a two players strategic game  differs from the SSE of the corresponding Stackelberg version of the game, meaning that the moves' order is crucial \cite{korzhyk2011stackelberg}.

\section{Results}\label{sec:result}

\subsection{Asymptotic Equilibrium and Emerging Network metrics}\label{sec:thm}

Let us define the risk measure of a node $\mathcal{R}_i$ as the probability that the node $i$ becomes infected, i.e. 
\begin{equation}\label{eq:risk1}
\mathcal{R}_i(\q,\bm{\phi};\bm{A}):= \mathbb{P}_i(\q,\bm{\phi};\bm{A}),
\end{equation}
where
\begin{eqnarray}\label{eq:P_i}
   \mathbb{P}_i(\q,\bm{\phi};\bm{A})&:=& \mathbb{E}_{\bm{X},s}\left[ \mathbb{I}\left ( i\sim s \in \mathcal{T}(\bm{X})\right) \right]\nonumber \\
   &=& \sum_{s=1}^n \phi_s \ \mathbb{E}_{\bm{X}}\left[ \mathbb{I}\left ( i\sim s \in \mathcal{T}(\bm{X})\right) \right], 
\end{eqnarray}
and $\mathbb{I}(\cdot)$ denotes the indicator function of a set. The expression $i \sim s \in \mathcal{T}(\bm{X})$ represents the event that node $i$ is connected to the seed node $s$ within the transmission subnetwork of susceptible nodes $\mathcal{T}(\bm{X})$, as in Definition~\ref{def:infection}.

The following theorem states that the SSE exists and it is unique in our framework.

\begin{theorem}[Existence and uniqueness of SSE]\label{thm:SSE}
    Consider a security Stackelberg game with contagion on a network $\mathcal{G}$ of $n$ nodes, defined by the attacker utility $\mathcal{U}_a$~\eqref{eq:utility}, the defender loss $\mathcal{L}_d$~\eqref{eq:loss}, a risk profile $\bm{\mathcal{R}}$ as in~\eqref{eq:risk1}, and quadratic cost functions. If $\alpha$ and $\theta$ are sufficiently large then there exists a unique SSE $(\q^*,\bm{\phi}^*(\cdot))$.
\end{theorem}

\begin{proof}
    The attacker's utility can be written in the following form: 
    $$
         \mathcal U_{a}(\bm{\phi};\bm{q})= {\bm w}^T \bm\phi- \frac{\theta}{2}\|\bm\phi\|^2, 
    $$
where $w_s=\sum_{i=1}^n \eta_i \mathbb{E}_{\bm{X}}\left[ \mathbb{I}\left ( i\sim s \in \mathcal{T}(\bm{X})\right) \right]$, therefore the Hessian matrix of $\mathcal U_a$ is $\nabla_{\bm \phi}^2\mathcal U_{a}(\bm{\phi};\bm{q})=-\theta I$ that is negative definite since $\theta>0$, so $\mathcal U_a$ is strictly concave in $\bm \phi$ for any $\bm q$. Taking into account that $S_a$ is compact and convex, we have that there exists a unique attacker's best response $\bm \phi^*(\bm q)=\text{argmax}_{\bm\phi}\mathcal U_a(\bm \phi, \bm q)$ for any $\bm q$.
For sufficiently large $\theta$, the quadratic term dominates ensuring that the optimum lies in the interior of $S_a$ so that the function $\bm q\mapsto \bm\phi^*(\bm q)$ is smooth.

Now, let us define $\tilde {\mathcal L}_d(\bm q):=\mathcal L_d(\bm q;\bm\phi^*(\bm q))$. Since $\mathcal R_i$ is a polynomial function of $\bm q$ and $\bm \phi$ as provided by \eqref{def:p(s,X)} and \eqref{eq:P_i}, and $\bm\phi^*$ is a smooth function of $\bm q$ then $\tilde{\mathcal L}_d$ is a smooth function of $\bm q$. Since $S_d$ is compact, there exists $\bm q^*=\text{argmin}_{\bm q} \tilde{\mathcal L_d}(\bm q)$. 

The Hessian matrix of $\tilde{\mathcal L_d}$ is: 
$$
\nabla^2 \tilde{\mathcal L_d}(\bm q)=\sum_{i=1}^n z_i\nabla^2\tilde{\mathcal R_i}(\bm q)+\alpha I,
$$
where $\tilde{\mathcal R_i}:\bm q\mapsto \mathcal R_i(\bm q, \bm\phi^*(\bm q))$. Since $S_d$ is compact and $\tilde{\mathcal R_i}$ is a smooth function then $\sum_{i=1}^n z_i\nabla^2\tilde{\mathcal R_i}$ is uniformly bounded. Therefore, for a sufficiently large $\alpha>0$, $\nabla^2\tilde{\mathcal L_d}$ is positive definite and $\tilde{\mathcal L_d}$ is strictly convex, implying the uniqueness of $\bm q^*$. 
\end{proof}

Once the existence and uniqueness of the SSE have been established, the next step is its characterization. In the considered context, we prove that the optimal security investment $\bm q^*$ corresponding to the SSE of the security game can be expressed using simple network metrics, which describe each node's vulnerability based on its position within the network topology.

To make it explicit, let us define two quantities called  $1$-point and $2$-points network protection metrics as follows.
\begin{definition}[$1$-point protection]\label{def:a}
Let $\mathcal{G}$ be a network of $n$ nodes. The $1$-point protection is defined as the tensor $\bm{p}^1 = \{a^j_{ik}\}_{i,j,k=1}^n \in \{0,1\}^{n^3}$, where each entry $a^j_{ik}$ is given by
$$
a^j_{ik} =
\begin{cases}
1 & \text{if } i\sim k \text{ in }\mathcal G \text{ and } i \not\sim k \text{ in } \mathcal{G} \setminus \{j\}, \\
0 & \text{otherwise},
\end{cases}
$$
and $\mathcal{G} \setminus \{j\}$ denotes the sub-network obtained by removing node $j$ from $\mathcal{G}$. We define $a_{ii}^j=0$, $\forall i\neq j$, and $a_{ij}^j=1$.
\end{definition}
The $1$-point protection encodes whether the removal of a single node $j$ can disrupt the connectivity between a pair of nodes $(i,k)$, initially connected, potentially blocking the spread of an infection between them. For example, in the network shown in Fig.~\ref{fig:protection}, the potential contagion between nodes $3$ and $6$ can be prevented by removing (or immunizing) either node $5$, so that $a^5_{36} = 1$, or node $1$, resulting in $a^1_{36} = 1$. From the $1$-point protection tensor, given two vectors $\bm{v},\bm{w}\in\mathbb{R}^n$, we define its weighted reductions  as
\begin{align}
a_i^j(\bm{v})&:= \sum_{k=1}^n a_{ik}^j v_k = \sum_{k=1}^n a_{ki}^j v_k \label{eq:a^j_i}\\
a^i(\bm{v},\bm{w})&:= \sum_{j,k=1}^n a_{jk}^i v_j w_k= \sum_{k}^n a_{k}^i(\bm{v})  w_k=\sum_{j}^n a_{j}^i(\bm{w})  v_j,\label{eq:a^i}
\end{align}
where we used that $a^j_{ik}=a^j_{ki}$ that also implies $a^i(\bm{v},\bm{w})=a^i(\bm{w},\bm{v})$. 

\begin{definition}[$2$-points protection]\label{def:b}
    Let $\mathcal{G}$ be a network of $n$ nodes. The $2$-point protection is defined as the tensor $\bm{p}^2 = \{b^{(i,j)}_{ks}\}_{i,j,k,t=1}^n \in \{0,1\}^{n^4}$, where each entry $b^{(i,j)}_{ks}$ is given by
$$
b^{(i,j)}_{ks} =
\begin{cases}
1 & \text{if}\ k \sim s \text{ in } \mathcal{G} \ , \ a^i_{ks}=a^j_{ks}=0
\ ,\ \ k \not\sim s \text{ in } \mathcal{G} \setminus \{i,j\}, \\
0 & \text{otherwise},
\end{cases}
$$
where $a^j_{ks}$ denotes the entries of the $1$-point protection tensor $\bm{p}^1$ of $\mathcal{G}$,  and $\mathcal{G} \setminus \{i,j\}$ is the sub-network obtained by removing both nodes $i$ and $j$ from $\mathcal{G}$. 
\end{definition}
Unlike the $1$-point protection, the $2$-point protection captures whether the simultaneous removal of a pair of nodes $(i,j)$ can disrupt the connectivity between another pair of nodes $(k,s)$, initially connected, in cases where the removal of either $i$ or $j$ alone is insufficient to cause such disruption.  For example, in the network shown in Fig.~\ref{fig:protection}, the removal (or immunization) of either node $2$ or node $4$ alone is not sufficient to prevent the potential contagion between nodes $3$ and $6$—that is, $a^2_{36} = a^4_{36} = 0$. However, the simultaneous removal of both nodes $(2,4)$ successfully blocks the connection, and thus $b^{(2,4)}_{36} = 1$. For a given pair of vectors $\bm{v},\bm{w}\in\mathbb{R}^n$, we define
\begin{equation}\label{eq:b_ij}
b_{ij}(\bm{v},\bm{w}):= (1-\delta_{ij}) \sum_{k,s} \left(b^{(i,j)}_{ks}  -a^i_{ks}a^j_{ks}  \right) v_kw_s.
\end{equation}

\begin{figure}
\centering
\begin{tikzpicture}[scale=0.7]

\node[circle, draw, fill=orange!50,label={[label distance=1mm]below right:1}] (n1) at (0,-2) {};
\node[circle, draw, fill=orange!50,label={[label distance=1mm]above left:2}] (n2) at (-1,0.5) {};
\node[circle, draw, fill=orange!50,label={[label distance=1mm]below left:3}] (n3) at (-2,-3) {};
\node[circle, draw, fill=orange!50,label={[label distance=1mm]above right:4}] (n4) at (2,-0.5) {};
\node[circle, draw, fill=blue!20,label={[label distance=1mm]above right:5},inner sep=0pt] (n5) at (0.5,1.5) {$\bm{\times}$};
\node[circle, draw, fill=red!50,label={[label distance=1mm]above left:6}] (n6) at (-0.,3.5) {\faBolt};

\draw[thick] (n1)--(n2);
\draw[thick] (n2)--(n5);
\draw[thick] (n4)--(n5);
\draw[thick] (n1)--(n3);
\draw[thick] (n1)--(n4);
\draw[thick] (n5)--(n6);


\begin{scope}[xshift=7cm]
\node[circle, draw, fill=orange!50,label={[label distance=1mm]below right:1}] (n1) at (0,-2) {};
\node[circle, draw, fill=blue!20,label={[label distance=1mm]above left:2},inner sep=0pt] (n2) at (-1,0.5) {$\times$};
\node[circle, draw, fill=orange!50,label={[label distance=1mm]below left:3}] (n3) at (-2,-3) {};
\node[circle, draw, fill=blue!20,label={[label distance=1mm]above right:4},inner sep=0pt] (n4) at (2,-0.5) {$\times$};
\node[circle, draw, fill=orange!50,label={[label distance=1mm]above right:5}] (n5) at (0.5,1.5) {};
\node[circle, draw, fill=red!50,label={[label distance=1mm]above left:6}] (n6) at (-0.,3.5) {\faBolt};

\draw[thick] (n1)--(n2);
\draw[thick] (n2)--(n5);
\draw[thick] (n4)--(n5);
\draw[thick] (n1)--(n3);
\draw[thick] (n1)--(n4);
\draw[thick] (n5)--(n6);

\draw[dashed] (-2.5,1) -- (3.5,-1);

\end{scope}

\end{tikzpicture}

\caption{Illustration of $1$- and $2$-point protection in a network. The potential contagion between nodes $3$ and $6$ can be prevented by removing (or immunizing) either node $5$ (left panel), so that $a^5_{36} = 1$, or node $1$, resulting in $a^1_{36} = 1$. In contrast, the removal of either node $2$ or node $4$ alone is not sufficient to block the contagion—i.e., $a^2_{36} = a^4_{36} = 0$—but their simultaneous removal (right panel) successfully disrupts the connection, yielding $b^{(2,4)}_{36} = 1$.  \label{fig:protection}}
\end{figure}

The $1$- and $2$-point protection metrics identify the most strategic and cost-effective locations in the network where a potential infection can be locally blocked. Since they rely on the removal (or immunization) of only one or two nodes, these metrics suggest defense strategies that are both targeted and resource-efficient. As such, they provide actionable guidance on where to concentrate cyber defense efforts and allocate investments.

Many other approaches exist to assess node importance in terms of system vulnerability and contagion prevention. For instance the previous framework could be generalized to define $k$-point protection for arbitrary values of $k$, allowing for the study of more complex cooperative defense strategies involving multiple nodes.
Moreover, centrality-based measures—such as degree, betweenness, closeness, and k-core—have long been used to identify influential nodes, and recent work has proposed multi-attribute decision-making methods that integrate several structural indicators to provide a more comprehensive assessment \cite{zhang2022multi}. What makes the $1$- and $2$-point protection metrics particularly relevant is that they emerge naturally from the analysis of Stackelberg equilibria in the security game introduced in this work. In other words, the relevant network metrics are not postulated ex ante, but emerge
endogenously from the equilibrium characterization. In fact, it holds the following 

\begin{figure}[h]
    \centering
    \includegraphics[width=0.495\textwidth]{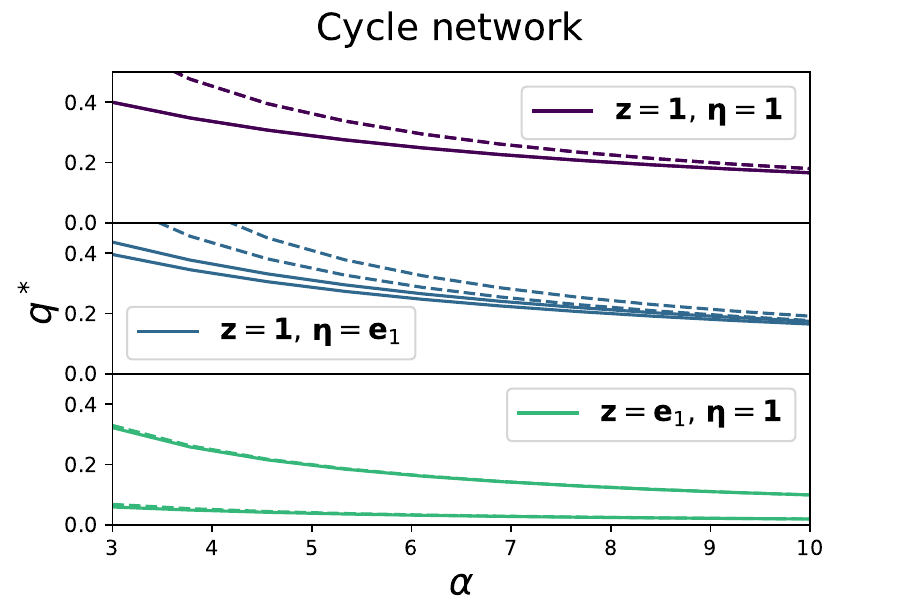}
    \includegraphics[width=0.495\textwidth]{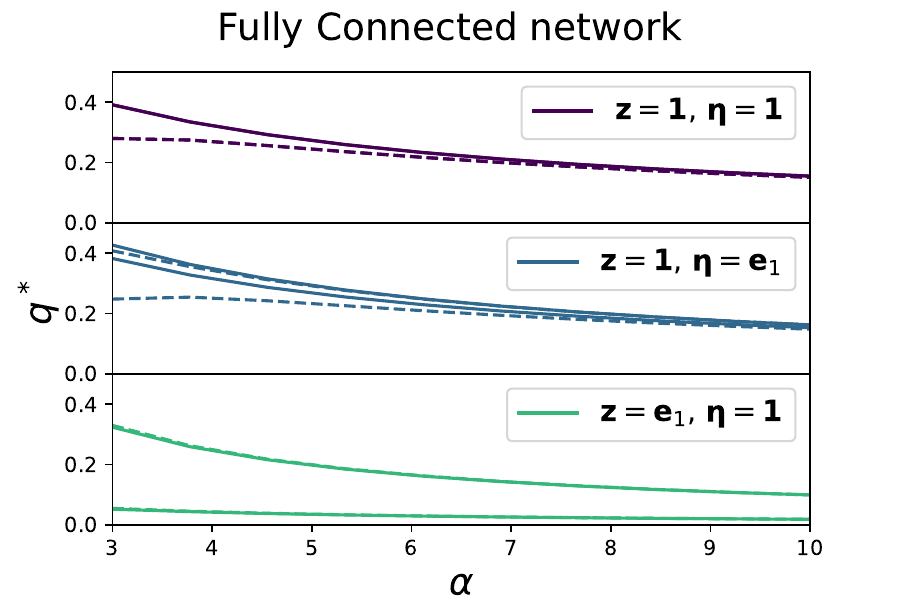}

    \caption{Exact (solid lines) and approximated (dashed lines) optimal investment strategies given by Theorem \ref{thm:approxSSE} for two symmetric network topologies: a ring network, where nodes are connected in a cycle (left panel), and a fully connected network (right panel). Results are shown for different choices of the value profiles $(\bm{z},\bm{\eta})$\label{fig:error}.}
\end{figure}

\begin{theorem}[Asymptotic SSE]\label{thm:approxSSE}
    Consider a security Stackelberg game with contagion on a network $\mathcal{G}$ of $n$ nodes, defined by the attacker utility $\mathcal{U}_a$~\eqref{eq:utility}, the defender loss $\mathcal{L}_d$~\eqref{eq:loss}, a risk profile $\bm{\mathcal{R}}$ as in~\eqref{eq:risk1}, and quadratic cost functions. If $\theta$ is sufficiently large, a SSE security investment satisfies
    \begin{equation}\label{eq:qstar}
    \q^*\sim^\alpha \left( \alpha \mathbb{I}-\bm{M} \right)^{-1} \bm{s},
    \end{equation}
where $\bm{M} = \bm{M}(\bm{z}, \bm{\eta}; \bm{p}^1, \bm{p}^2) \in \mathbb{R}^{n \times n}$ and $\bm{s} = \bm{s}(\bm{z}, \bm{\eta}; \bm{p}^1, \bm{p}^2) \in \mathbb{R}^n$ are defined by 
\begin{align}\label{eq:a^ib_ij}
s^i(\bm{z}, \bm{\eta}; \bm{p}^1, \bm{p}^2)&:= a^i(\bm{1}/n,\bm{z}),\nonumber\\
M_{ij}(\bm{z}, \bm{\eta}; \bm{p}^1, \bm{p}^2)&:= b_{ij}(\bm{1}/n,\bm{z}) \nonumber \\
& -\frac{1}{\theta}\sum_{k=1}^n \left( a^i_k(\bm{\eta})a^j_k(\bm{z})  - a^i(\bm{1}/n,\bm{\eta})a^j(\bm{1}/n,\bm{z}) \right)\nonumber \\
 &-\frac{1}{\theta}\sum_{k=1}^n \left( a^i_k(\bm{z})a^j_k(\bm{\eta})  - a^i(\bm{1}/n,\bm{z})a^j(\bm{1}/n,\bm{\eta}) \right)
\end{align}
The notation $\bm{a} \sim^\alpha \bm{b}$ indicates that $\|\bm{a} - \bm{b}\| = o(1/\alpha^2)$ as $\alpha\to\infty$.
\end{theorem}
The proof of Theorem~\ref{thm:approxSSE} is provided in Appendix~\ref{sec:app} and builds upon a generalization of the results presented in~\cite{acemoglu2016network}. Theorem~\ref{thm:approxSSE} offers an explicit approximation of the defender’s optimal cyber investment strategy, which becomes asymptotically exact in the limit $\alpha \to \infty$. This corresponds to a scenario in which the cost function dominates the optimization, and can therefore be interpreted as a low-budget regime for the defender. In fact, Eq. (\ref{eq:qstar}) can be expanded as \begin{equation}\label{eq:qstarexpanded}
\q^*= \frac{1}{\alpha}\bm{s} +\frac{1}{\alpha^2}\bm{M}\bm{s} +o\left(\frac{1}{\alpha^2}\right),
\end{equation}
indicating a vanishing optimal investment as $\alpha\to\infty$. Fig. \ref{fig:error} illustrates the agreement between the exact and approximated optimal investment strategies across two network topologies and multiple value profile choices.

A key feature of this approximation is its dependence on the network topology solely through the network-based metric tensors $\bm{p}^1$ and $\bm{p}^2$, which naturally emerge from the structure of the game and are combined properly according to the value profiles $\bm{z}$ and $\bm{\eta}$. From an economic perspective, Eq. \ref{eq:qstar} shows that efficient cybersecurity investments are jointly determined by network externalities and heterogeneous asset valuations. As a result, optimal resource allocation cannot be inferred solely from topological characteristics such as connectivity or centrality, but must also account for the economic importance that attackers and defenders assign to different components of the system. This interaction between network structure and heterogeneous risk perceptions represents the main mechanism through which cyber-risk heterogeneity shapes equilibrium behavior.

Theorem \ref{thm:approxSSE} extends the results of \cite{acemoglu2016network}, which are recovered as special cases. For instance, in the limit as $\theta \to \infty$, the utility function reduces to the cost $\mathcal{C}_a(\bm{\phi})$, compelling the attacker to adopt a uniform strategy $\bm{\phi}^* = \bm{1}/n = (1/n, \ldots, 1/n)$. In this regime, the problem simplifies to the social optimum under random, non-strategic attacks, as studied in \cite{acemoglu2016network} for $\bm{z} = \bm{1}$.
Another special case arises by setting $\bm{z} = \bm{e}_i$ and $\bm{\eta} = \bm{1}/n$. Under these conditions, we obtain $a^i_k(\bm{z}) = a^i(\bm{1}/n, \bm{z}) = 1$, and $b_{ij}(\bm{1}/n, \bm{z}) = -(1 - \delta_{ij}) a^j_i(\bm{1}/n)$. Consequently, the matrix $M$ becomes
$$M_{ij}=-(1-\delta_{ij})a^j_i(\bm{1}/n) -\frac{1}{\theta}\sum_k a^i_k(\bm{1}/n)(a^j_{ik}-a^j_i(\bm{1}/n)),$$
which characterizes the Nash equilibrium under uniform strategic attack, as analyzed in \cite{acemoglu2016network}, where each node in the network acts as an individual player minimizing its own infection probability. Again in the limit as $\theta \to \infty$ we recover the Nash equilibrium under random uniform attack.

Moreover, the statement of Theorem \ref{thm:approxSSE} reveals new economically meaningful mechanisms linking heterogeneity of the model, network structure, and strategic behavior. A first mechanism can be interepreted as a {\em heterogenity amplification}. The attacker's and defender's preferences affect the equilibrium allocation through $\bm s$ and $\bm M$, defined in \eqref{eq:a^ib_ij}, which arise as contractions of the tensors $p_1$ and $p_2$ weighted by the profiles $\bm z$ and $\bm\eta$. As a results, the map $(\bm z,\bm \eta)\mapsto \bm q^*(\bm z, \bm \eta)$ is nonlinear, and small perturbation in the value profiles may generate large shifts in equilibrium investments. To make this intuition more formal, consider a perturbation of the profiles $(\bm z,\bm \eta)$ and denote by $\delta \bm s$, $\delta \bm M$, and $\delta \bm q^*$ the corresponding variations. By expansion \eqref{eq:qstar}, we simply get: 
\begin{equation}
    \delta \bm q^*=\frac{1}{\alpha}\delta\bm s+\frac{1}{\alpha^2}\left(\bm M\delta \bm s+(\delta\bm M)\bm s\right)+o(\alpha^{-2}).
\end{equation}
 Therefore, whenever $\bm M$ has large spectral norm, even small perturbations in $\bm z$ and $\bm \eta$ can induce large changes in the equilibrium allocation $\bm q^*$.

The spectral norm of $\bm M$ tends to be large in the presence of structural asymmetries in the network. In particular, consider networks with bottlenecks or highly central nodes: if a small subset of nodes lies on a large fraction of contagion paths, then the corresponding entries of the tensors $p_1$ and $p_2$ become highly concentrated. Through the contractions defining $\bm M$, this concentration translates into large rows or columns of $\bm M$, and therefore into a large spectral norm. Intuitively, when contagion paths or strategic effects are concentrated on a few nodes, changes affecting those nodes propagate strongly across the system. This generates a strong amplification of the impact of preference heterogeneity on equilibrium investments.

A second mechanism arises from the sequential Stackelberg structure of the game and can be interpreted as a strategic {\em defensive misallocation} effect w.r.t. the preferences provided by the proposed heterogeneous model. Since the defender moves first and anticipates the attacker's best response, optimal investments decisions are not driven solely by the node preferences $z_i$, but also by each node's strategic role in shaping the attacker's behavior and the resulting attack distribution. In the heterogeneous setting, some nodes are over-protected not because they are intrinsically the most valuable, but because they act as strategic leverage points that redirect the attacker's response or as critical intermediaries in contagion pathways. This intuitive effect is captured by the matrix $\bm M$ defined in \eqref{eq:a^ib_ij}, whose off-diagonal entries describe how protecting one node changes the marginal value of protecting others through contagion externalities. Consequently, the presence of heterogeneity and sequentiality allow non-trivial strategic allocations.

From a mathematical viewpoint, misallocation w.r.t. the defender's preferences is driven by the matrix $\bm M$. In fact, in the absence of its contribution, the equilibrium reduces to $\bm q^*=\alpha^{-1}\bm s$ by \eqref{eq:qstarexpanded}. Since $s^i=\frac{1}{n}\sum_{j,k} a^i_{jk} z_k$,  
in this case the defenders invests mainly on the nodes which serve as protections of other nodes that are important according to the preferences $\bm z$. In this sense, the defense allocation is coherent to the defender's preferences. Misallocation, related to the sequentiality of the game, are introduces by the action of $\bm M$ and its dependency on $\bm\eta$. Therefore, the magnitude of defense misallocations is controlled by the spectral norm of $M$:
$$\left\|q^*-\frac{1}{\alpha}\bm s\right\| \leq\frac{1}{\alpha^2} \|M\|\|s\|.$$

Within the proposed heterogenous model, we have that economic mechanisms such as heterogeneity amplification and defense misallocations are induced by asymmetric network structures that are mathematically captured by the spectral norm of the matrix $\bm M$ introduced in \eqref{eq:a^ib_ij}.

\subsection{Paths of contagion and scalable solutions}\label{sec:path}
The analytical expression provided by Theorem \ref{thm:approxSSE} offers a good approximation of the optimal investment in the small-budget regime. Nevertheless, it can serve as a benchmark strategy across a broader range of scenarios. Its main advantage lies in its computational simplicity, as it avoids solving the full optimization problem, which typically requires a costly numerical search for the true SSE.

The core challenge lies in computing the risk profile $\bm{\mathcal{R}}$ as in ~\eqref{eq:risk1}, which in turn determines the utility and loss functions to be optimized. Introducing the random variable
\begin{align}
N^L_{i}(s,\bm{X};\bm{A}) = \mathrm{Card}&\left( \left\{ (i_1 = i, i_2, \ldots, i_\ell = s)  \,\middle|\, A_{i_k i_{k+1}} = 1,\; X_{i_k} = 1,\; \right.\right.\nonumber \\
&\quad \left. \left.  i_k \in \mathbb{N}_n,\; k \in \mathbb{N}_{\ell-1},\; \ell \leq L \right\}  \right)
\end{align}
as the number of different paths of maximum length $L$ connecting a node $i$ to the seed $s$ in the transmission subnetwork $\mathcal{T}(\bm{X})$ of susceptible nodes, the infection probability can be written as
\begin{align}\label{eq:frisk}
   \mathbb{P}_i(\q,\bm{\phi};\bm{A})
   &=\mathbb{E}_{\bm{X},s}\left[ \mathbb{I}\left ( i\sim s \in \mathcal{T}(\bm{X})\right) \right]\nonumber  \\
   &= \mathbb{E}_{\bm{X},s}\left[ \lim_{L\to\infty}\Theta\left(N^L_i(s,\bm{X};\bm{A}) -1\right) \right], \nonumber  \\
   &=\lim_{L\to\infty}\mathbb{E}_{\bm{X},s}\left[ \Theta\left(N^L_i(s,\bm{X};\bm{A}) -1\right) \right],
\end{align}
where $\Theta(\cdot)$ denotes the Heaviside step function which returns one if the argument is greater than zero, i.e. $N^L_i \geq 1$, and zero otherwise. In the second line of Eq. \ref{eq:frisk} we exploit the fact that, for a given $i\in\mathbb{N}_n$, the sequence of random variables $(\Theta(N^L_i-1))_{L\geq 0}$ converges pointwise to $\mathbb{I}\left ( i\sim s \in \mathcal{T}(\bm{X})\right)$. Intuitively, a node $i$ is connected to the seed in $\mathcal{T}(\bm{X})$ if there exists at least a path, of arbitrary large length $L$, linking them. In the last line, the application of the dominated convergence theorem justifies the exchange of limit and expectation. 
Following the definition of $N^L_{i}(s,\bm{X};\bm{A})$, let us observe that the parameter $L$ is not only a technical cut-off but admits a natural interpretation as propagation time. In fact, a path of length $L$ can be regarded as a sequence of $L$ successive transmissions of the infection through the network to which we can associate a unit of time. In this sense, the path length directly corresponds to the time required for the contagion to travel from the seed to another node. Considering paths of length at most $L$ is therefore equivalent to assuming that the infection can propagate only within a finite time interval. While, in the limit $L\rightarrow \infty$, the model recovers the regime in which the contagion has unlimited time to diffuse. In this sense, $L$ can be understood as the infection propagation time, which equips the model with an explicit dynamical dimension.

The form of Eq. \ref{eq:frisk} suggests the introduction of an entire class of risk measure defined in terms of a pair $(f,L)$ as
\begin{align}\label{eq:genfrisk}
   \mathcal{R}^{(f,L)}_i(\q,\bm{\phi};\bm{A})
   &=\mathbb{E}_{\bm{X},s}\left[ f\left(N^L_i(s,\bm{X};\bm{A}) \right) \right],
\end{align}
where $f$ is an arbitrary non-decreasing activation function. If $f$ is bounded one can also consider the limit $\mathcal{R}_i^f=\lim_{L\to\infty} \mathcal{R}_i^{(f,L)}$.

In general, if the activation function $f$ is non linear, the expected value cannot be computed explicitly and must instead be evaluated numerically. The computational cost of this operation grows exponentially with the network size $n$. Conversely, if we consider a linear activation function, then the risk is expressed as the expected number of paths from the node to the seed, i.e.
\begin{align}
    \mathcal{R}^{L}_i(\q,\bm{\phi};\bm{A})
   &=\mathbb{E}_{\bm{X},s}\left[ N^L_i(s,\bm{X};\bm{A}) \right]= \mathbb{E}_{\bm{X},s}\left[\sum_{\ell=1}^L  \left( \bm{A} \circ \bm{X}\bm{X}^T \right)^{\ell}_{is}  \right]\label{eq:riskpath} \\
   &=\mathbb{E}_{\bm{X},s}\left[\sum_{\ell=1}^L  \sum_{i_2,\ldots,i_{\ell -1}=1}^n A_{ii_2}A_{i_2i_3}\cdots A_{i_{\ell-1}s} X_iX_{i_2}\cdots  X_{i_{\ell-1}}X_s  \right]\nonumber\\
   &=\sum_{\ell=1}^L  \sum_{i_2,\ldots,i_{\ell -1},s=1}^n \phi_s A_{ii_2}A_{i_2i_3}\cdots A_{i_{\ell-1}s} \prod_{k\in\{i,i_2,\ldots,i_{\ell-1,s}\}^{(1)}} (1-q_k), \nonumber 
\end{align}
where we denoted with $I^{(1)}$ the set of distinct occurrences of a set $I$.
This quantity can be computed as efficiently as a matrix multiplication. Empirically, we observe that the results exhibit very low sensitivity to the cut-off parameter $L$. Therefore one can simply solve the optimization problem with $ \bm{\mathcal{R}}=\bm{\mathcal{R}}^{L}$ with very small $L$ to obtain optimal strategies that scale efficiently to large systems.

\section{Numerical results}\label{sec:num}

\subsection{Efficient frontier}\label{sec:effront}

For any given defender and attacker value profiles $\bm{z}$ and $\bm{\eta}$, the optimal defender strategy at the Stackelberg equilibrium provides a quantitative measure of the vulnerability associated with a specific network topology. This insight can be leveraged to guide the selection of system architectures that offer greater robustness. 

Inspired by the classical Markowitz portfolio selection theory \cite{markowitz1952portfolio}, we assess the robustness of different network topologies by computing the efficient frontier of the problem. Specifically, we scatter plot the system's global risk at equilibrium,
\begin{equation}
\mathcal{R}^*(\alpha;\bm{A}) := \sum_{i=1}^n z_i\, \mathcal{R}_i\left(\q^*(\alpha;\bm{A}),\, \phi^\star\left(\q^*(\alpha;\bm{A})\right);\, \bm{A}\right),
\end{equation}
highlighting its dependence on the network topology $\bm{A}$, against the corresponding defender's cost at equilibrium,
\begin{equation}
\mathcal{C}^*_d(\alpha;\bm{A}) := \mathcal{C}_d\left(\q^*(\alpha;\bm{A})\right),
\end{equation}
for varying values of the Lagrange multiplier $\alpha$. The resulting risk–cost plot illustrates the trade-off between investment and security: it shows the minimum cost the defender must incur to achieve a given level of system risk, or equivalently, the minimum achievable risk for a given admissible cost (i.e., the defender's budget). The security performance of different network topologies can thus be evaluated by comparing their respective efficient frontiers. The efficient frontier also serves as a powerful tool to compare the impact of different value profiles—that is, different attacker targeting regions and defender protection priorities—on a fixed system topology. By varying the allocation of value across nodes, we can assess how the equilibrium risk–cost trade-off shifts, providing insight into which regions of the network are most critical to secure under different threat scenarios.  

\begin{figure}[h]
    \centering
    \includegraphics[width=0.32\textwidth]{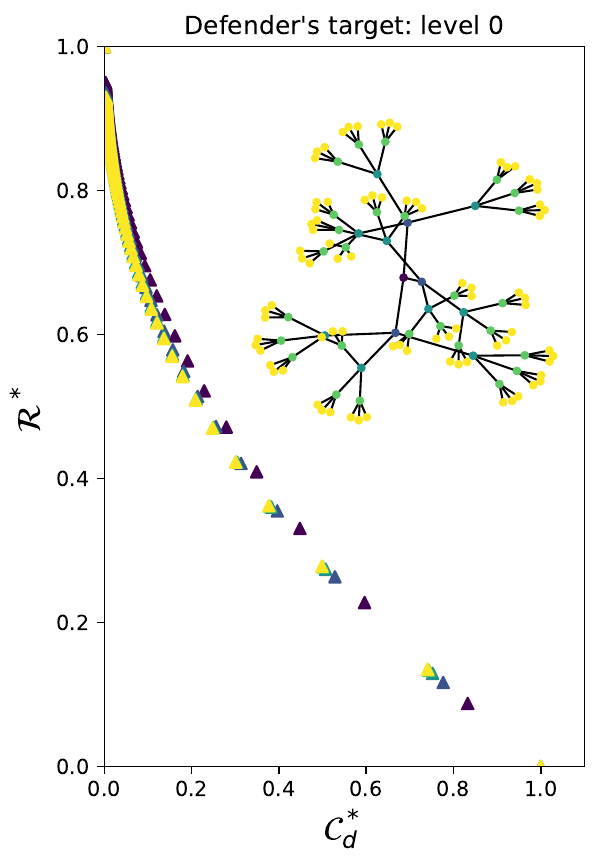}
    \includegraphics[width=0.32\textwidth]{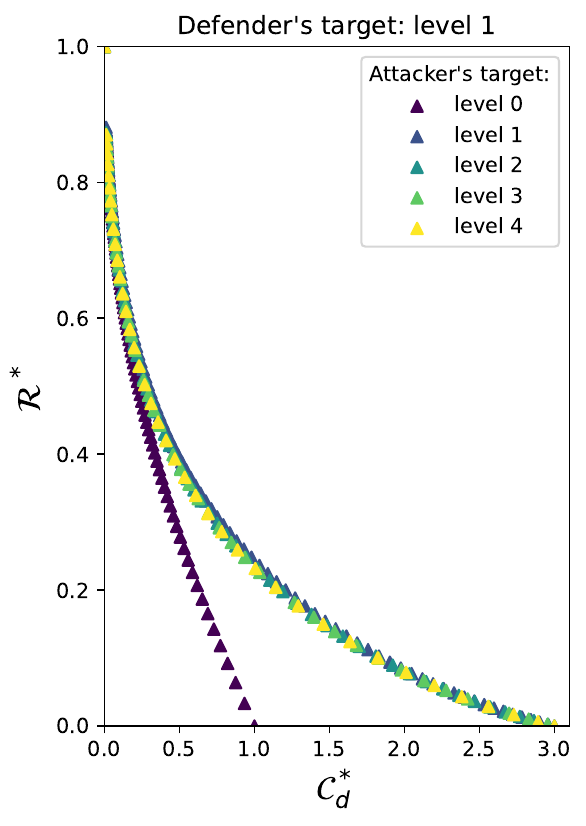}
    \includegraphics[width=0.32\textwidth]{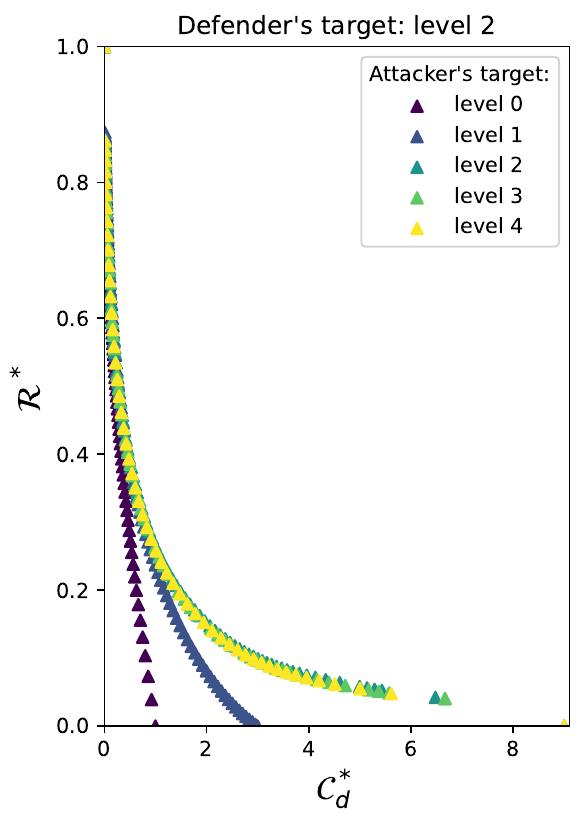}
    \caption{Efficient frontiers for a tree network with 121 nodes, branching ratio $3$, and $4$
 levels, are shown for different defender's value profile and attack distribution $\bm{z}$, $\bm{\phi}$ concentrated on specific levels of the tree. Colors distinguish the different levels targeted by the attacker (see inset of the Left Panel). Each panel corresponds to a different defender’s value profile, aiming to protect respectively the root (Left Panel), the first level (Central Panel), and the second level of the tree (Right Panel).\label{fig:effront_value}}
\end{figure}

Fig.~\ref{fig:effront_value} illustrates the efficient frontiers  for a tree-structured network consisting of $121$ nodes organized across four levels, with a branching ratio of $3$. The risk measure used is the infection probability (\ref{eq:risk1}) and costs are quadratic. We consider different value profiles $\bm{z}$ and  attack distribution $\bm{\phi}$, each targeting a distinct level of the tree—from the root, i.e. $\bm{z},\bm{\phi}= (1, 0, \ldots, 0)$, to intermediate levels such as $\bm{z},\bm{\phi}= (0, 1/3, 1/3,1/3, 0, \ldots, 0)$, and finally to the leaves, i.e. $\bm{z},\bm{\phi}= (0, \ldots, 1/81, \ldots, 1/81)$. One can observe that, it generally appears easier for the defender to protect levels that are closer to the root, as the efficiency frontiers tend to rise with the distance from the root in the defender’s value profile. However, the systemic risk is strongly affected by the attack distribution. In particular, the system appears less vulnerable when the attack start from inner levels than the one deemed most important by the defender. The reason is that a smaller subset of nodes must be immunized in order to prevent infection spread. In contrast, when the attack originates from peripheral levels, the risk vanishes once the defender’s budget is sufficient to immunize all nodes within the level of interest, regardless of the level targeted by the attacker. 

\begin{figure}[h]
    \centering
\includegraphics[width=0.5\linewidth]{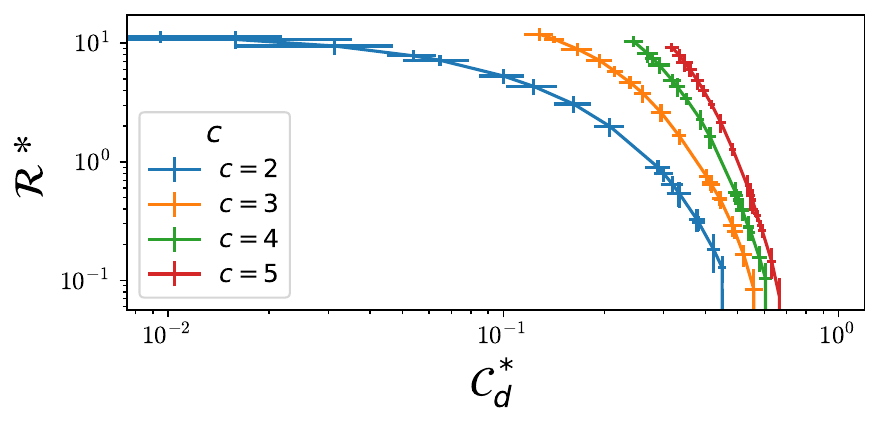} \caption{Efficient frontiers for Erdos-Renyi random networks with different connectivity. The  defender's cost is linear and the risk is $ \bm{\mathcal{R}}=\bm{\mathcal{R}}^{L=4}$. The value profiles $\bm{z}=\bm{\eta}=\bm{1}$ are uniform.}
    \label{fig:front_conn}
\end{figure}

Fig. ~\ref{fig:front_conn} shows the efficient frontiers of random networks, generated as instances of the sparse Erdős–Rényi random graph model with connectivity $c$, representing the average number of neighbors per node. In this experiment, we assume a linear defender’s cost and a risk $ \bm{\mathcal{R}}=\bm{\mathcal{R}}^{L}$ as in Eq. (\ref{eq:riskpath}) and the value profiles are uniform $\bm{z}=\bm{\eta}=\bm{1}$. We examine the effect of network topology by varying the connectivity level. Clearly, highly interconnected systems are both more vulnerable and harder to defend, as infections can spread along multiple pathways and are difficult to contain.
\begin{figure}[h]
    \centering
    \includegraphics[width=0.425\linewidth]{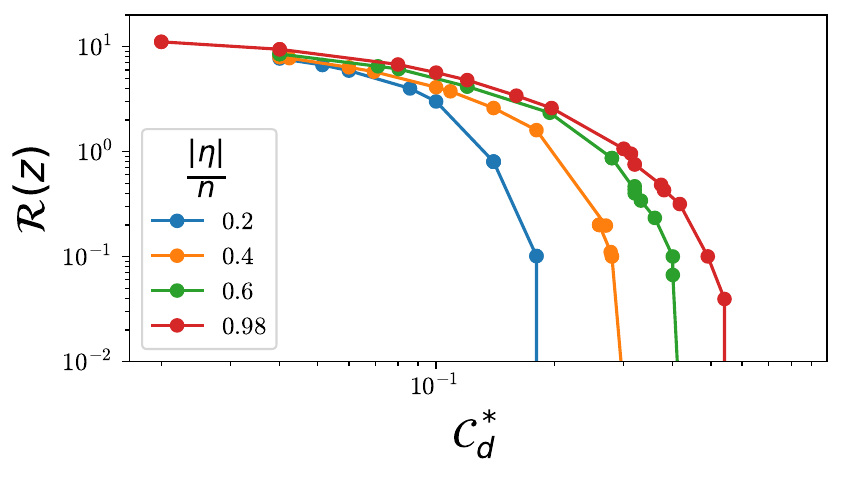}
\includegraphics[width=0.5\linewidth]{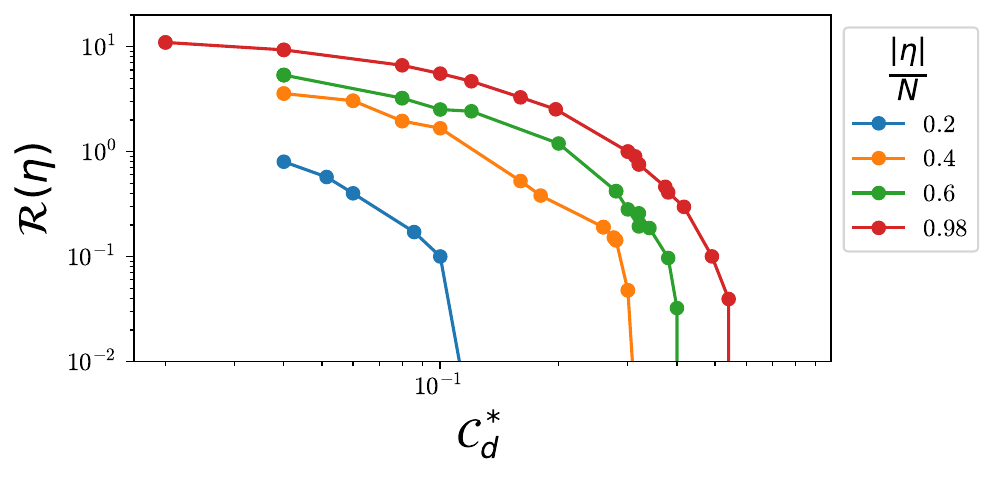}
    \caption{
    Efficient frontiers for an Erdos-Renyi random network, with linear defender's cost and risk $ \bm{\mathcal{R}}=\bm{\mathcal{R}}^{L=4}$. The defender's value profile $\bm{z}=\bm{1}$ is uniform while the attacker's value profile changes: $\bm{\eta}\in\{0,1\}^n$, $|\bm{\eta}|:=\sum_i \eta_i$. Left panel: optimal risk for the defender $\mathcal{R}^*(\bm{z})=\sum_i z_i \mathcal{R}^*_i $. Right Panel: optimal risk for the attacker $\mathcal{R}^*(\bm{\eta})=\sum_i \eta_i \mathcal{R}^*_i $
    }
    \label{fig:phasediagram}
\end{figure}

\begin{figure}[h]
    \centering
    \includegraphics[width=0.47\linewidth]{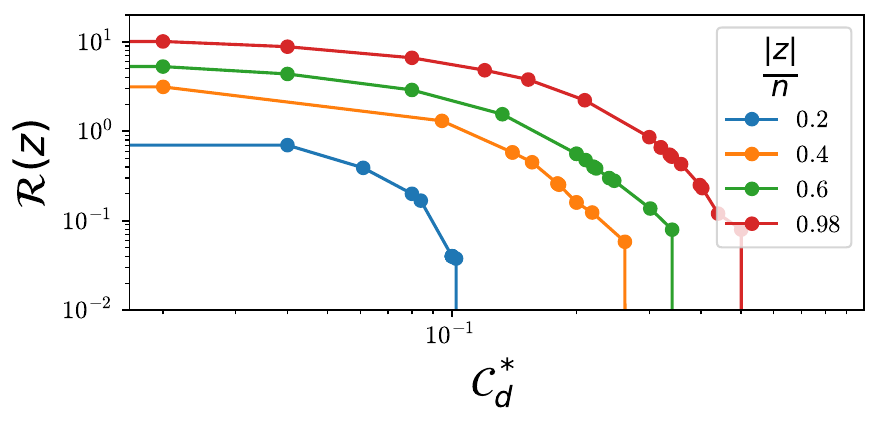}
\includegraphics[width=0.47\linewidth]{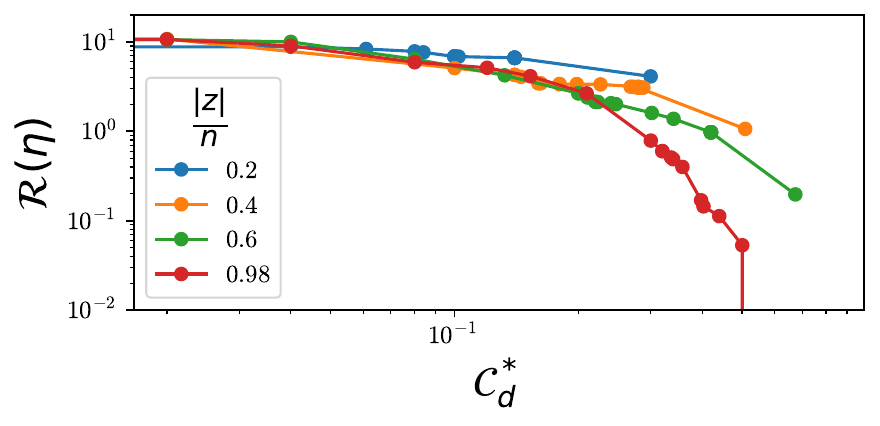}
    \caption{
    Efficient frontiers for an Erdos-Renyi random network, with linear defender's cost and risk $ \bm{\mathcal{R}}=\bm{\mathcal{R}}^{L=4}$. The attacker's value profile $\bm{\eta}=\bm{1}$ is uniform while the defender's value profile changes: $\bm{z}\in\{0,1\}^n$, $|\bm{z}|:=\sum_i z_i$. Left panel: optimal risk for the defender $\mathcal{R}^*(\bm{z})=\sum_i z_i \mathcal{R}^*_i $. Right Panel: optimal risk for the attacker $\mathcal{R}^*(\bm{\eta})=\sum_i \eta_i \mathcal{R}^*_i $ 
    }
    \label{fig:phasediagram2}
\end{figure}

In Fig. \ref{fig:phasediagram}, for a given Erdos-Renyi network and uniform defender value profile $\bm{z}=\bm{1}$, we investigate the effect of changing $\bm{\eta}$. In particular we choose $\bm{\eta}\in\{0,1\}^n$, for different  $|\bm{\eta}|:=\sum_i \eta_i$, measuring the size of the subset of nodes that are valuable for the attacker. We measure both the defender's risk $\mathcal{R}^*(\bm{z})=\sum_i z_i \mathcal{R}^*_i $ and the attacker's risk $\mathcal{R}^*(\bm{\eta})=\sum_i \eta_i \mathcal{R}^*_i $, meaning the risk associated with the specific portion of the network relevant to the different players. The results indicate that a more localized attacker value profile leads to a lower systemic risk. Furthermore, the phase transition to zero risk occurs at a smaller defender’s budget threshold.

Conversely, Fig. \ref{fig:phasediagram2} shows the opposite situation in which the  attacker  value profile $\bm{\eta}=\bm{1}$ is uniform while  the size of the subset of nodes that are valuable for the defender changes, i.e. $\bm{z}\in\{0,1\}^n$, for different $|\bm{z}|:=\sum_i z_i$.
Again it appears easier for the defender to pretect portion of the netowrk of smaller size, at a given budget. However the risk for the rest of the network increases, since the defensive strategy becomes suboptimal against systemic risk.

More generally, the efficient-frontier analysis shows that the effectiveness of cybersecurity spending depends not only on the amount of resources invested but also on how these resources are allocated across heterogeneous assets and network positions. As a result, systems facing similar cybersecurity budgets may achieve substantially different levels of resilience depending on the distribution of cyber risk and asset criticality.

\subsection{Cyber-deception  strategy effects} \label{sec:dec}

A detailed examination of the optimal defender’s strategy reveals how the investment is distributed across the various nodes of the network.
\begin{figure}[h]
    \centering
    \includegraphics[width=0.29\textwidth]{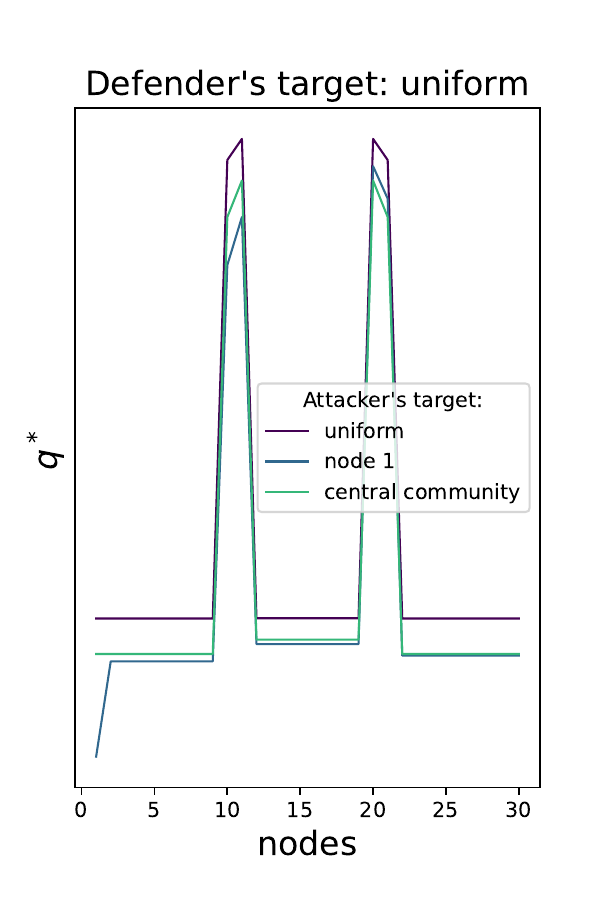}
    \includegraphics[width=0.29\textwidth]{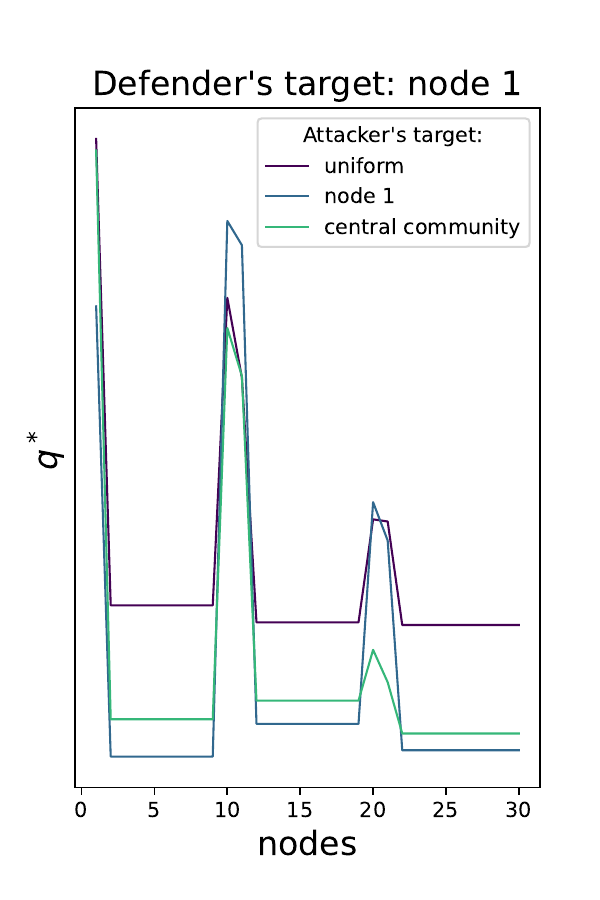}
    \includegraphics[width=0.325\textwidth]{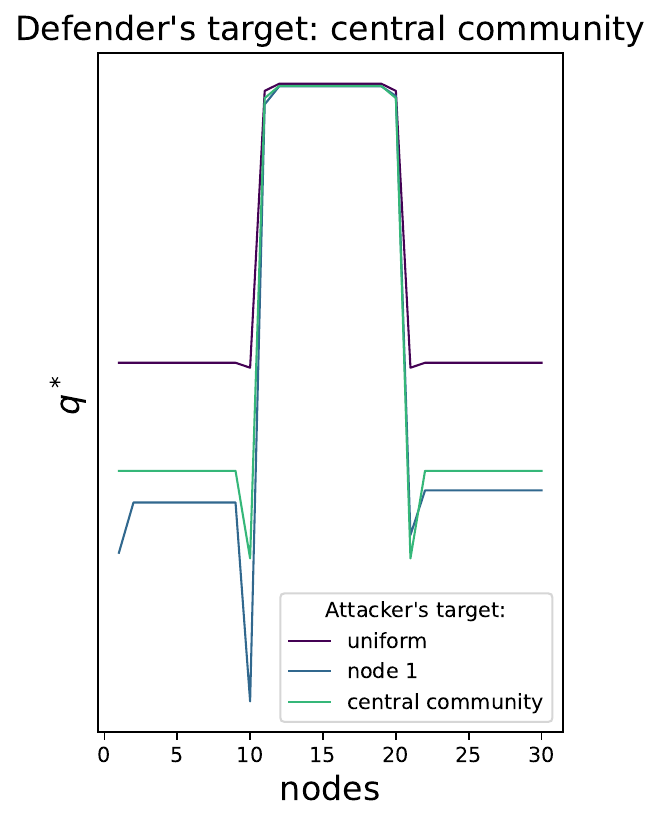}
    \caption{Optimal investment pattern $\bm{q}^*$ in a community network composed by three groups of 10 nodes, connected through the links ($10,11$) and $(20,21)$, for different value profiles $\bm{\eta}$ and $\bm{z}$.  Each panel corresponds to a
different defender’s value profile, aiming to protect respectively the whole system (Left Panel),
the first node (Central Panel), and the central community (Right Panel).  Different colors refers to different attacker's value profile.\label{fig:deception}}

    \end{figure}

In Fig. \ref{fig:deception} for example we display the optimal investment pattern, computed used the approximation of Theorem \ref{thm:approxSSE} through protection metrics,  in a network of $30$ nodes with assortative communities, for different value profile scenarios. Specifically, we examine all nine possible combinations of $(\bm{z},\bm{\eta})$, obtained by letting each value profile fall into one of the following three configurations: (i) all nodes in the network have identical value, i.e. $\bm{z},\bm{\eta}=(1/30, \ldots, 1/30)$; (ii) node $1$ is the only valuable node, namely $\bm{z},\bm{\eta}=(1, \ldots, 0)$; (iii) all nodes belonging to the central community share equal value, while all others have zero value, i.e. $\bm{z},\bm{\eta}=(0, \ldots,0, 1/10, \ldots,1/10, 0,\ldots, 0)$. Across the three plots, the defender’s value profile remains constant, whereas the attacker’s profile changes.

Nodes that serve as inter-community bridges tend to be the most strategic to protect, given their key role in halting the spread of infections when immunized. As expected, when the value profiles correspond to cases (i) or (iii), the investment pattern appears symmetric with respect to the central community. In contrast, this symmetry breaks when either $\bm{z}$ or $\bm{\eta}$ falls under case (ii). Furthermore, it is unsurprising that nodes in proximity to the defender’s most valuable assets attract greater security investment.

However, several unexpected effects become clearly observable. In particular, the attacker’s most valuable nodes consistently appear under-protected compared to others. Specifically, when node 1 becomes the most valuable for the attacker, its level of protection drops significantly (orange patterns); similarly, when the central community is the attacker’s primary target, its relative investment—compared to that of other communities—decreases (green pattern). This implies that the attacker’s optimal choice is not to launch a direct assault on the most valuable nodes. Rather, the attacker tends to initiate the infection from a seed located at some distance from them. Combined with the sequential nature of the Stackelberg game, this behaviour enables the attacker to mislead the defender, who ultimately disperses resources away from the truly critical assets. This type of effect arises when the two players assign different perceived values to the same node, while becomes much weaker when the attacker’s and defender’s value profiles are aligned.

 Instances of such asymmetry naturally emerge when the defender deploys a cyber‑deception protocol. We define \textit{deception} as any defender strategy that deliberately manipulates the attacker’s perception of the environment, i.e. the value, location, or accessibility of network assets, with the goal of inducing the attacker to behave suboptimally with respect to the real asset distribution. In our model, deception is the intentional introduction of a mismatch between the defender’s value profile $\mathbf{z}$ and the attacker’s perceived value profile $\bm{\eta}$.  This can be implemented for example by modifying the network structure itself. In practical cybersecurity settings, defenders routinely deploy decoy nodes (\textit{honeypots}) that imitate genuine services or machines, or  \textit{honeynets}, i.e. collections of interconnected decoy systems designed to emulate entire subnetworks. These artificial nodes do not contribute to the defender’s true value profile $\mathbf{z}$, but are intentionally crafted to appear valuable to the attacker, thereby altering the attacker’s perceived profile $\bm{\eta}$. By enlarging the apparent attack surface, providing false high‑value targets, or replicating critical components of the real infrastructure, structural deception aims to misdirect the attacker’s effort away from genuinely critical assets.

\begin{figure}[h]
    \centering
    \includegraphics[width=0.49\textwidth]{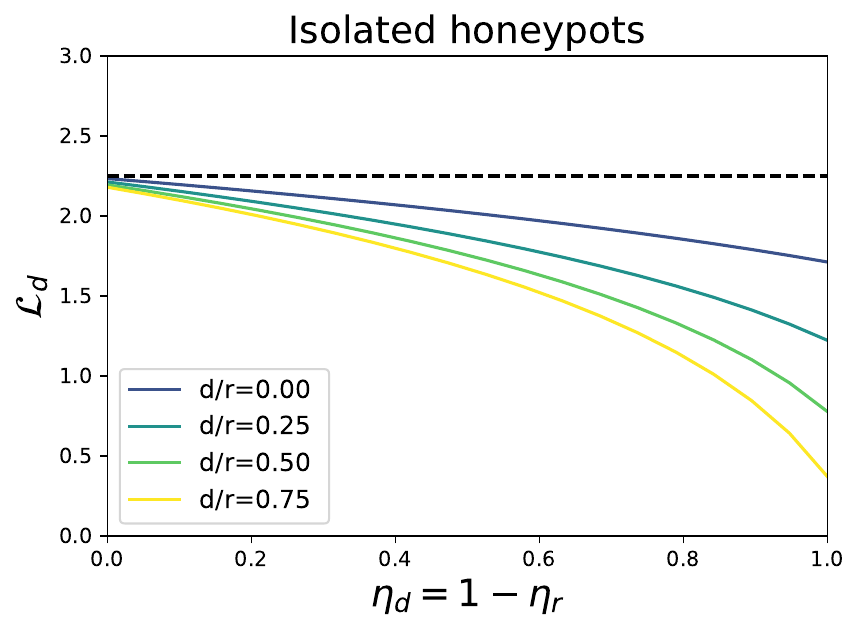}
    \includegraphics[width=0.49\textwidth]{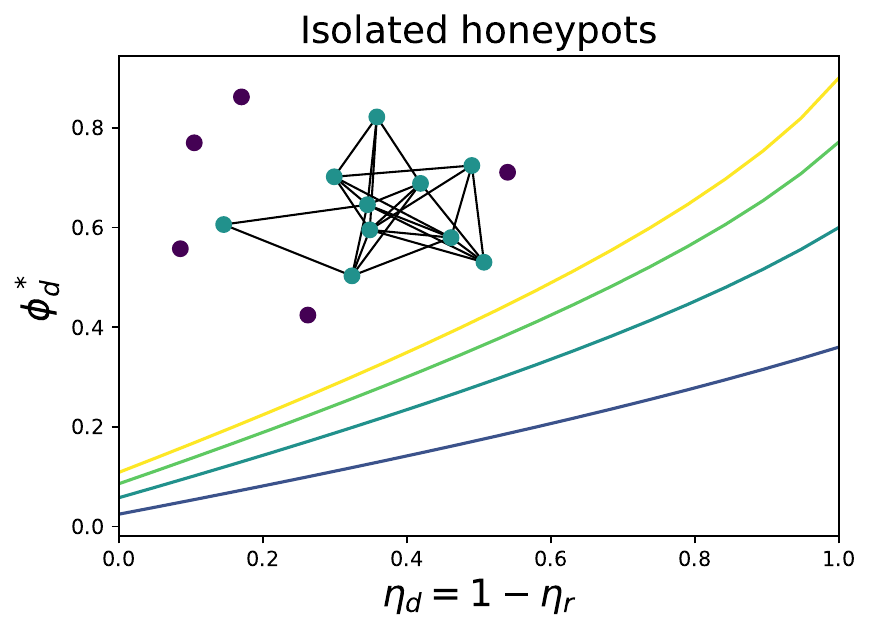}
    \caption{Cyber deception in a network with $r$ nodes and $d$ isolated honeypots with attacker perceived value $\eta_d=1-\eta_r\in[0,1]$. The deception level $\phi^*_d=\sum_{i\in N_d} \phi^*_i$ (right panel) and the effectiveness of the strategy (left panel) relative to the case without honeypots (dashed line) are shown for increasing fraction of honeypots. Since honeypots are isolated, the cost of deception is always negligible}.\label{fig:isolated_honeypoint}
    \end{figure}

\begin{figure}[h]
    \centering
    \includegraphics[width=0.3\textwidth]{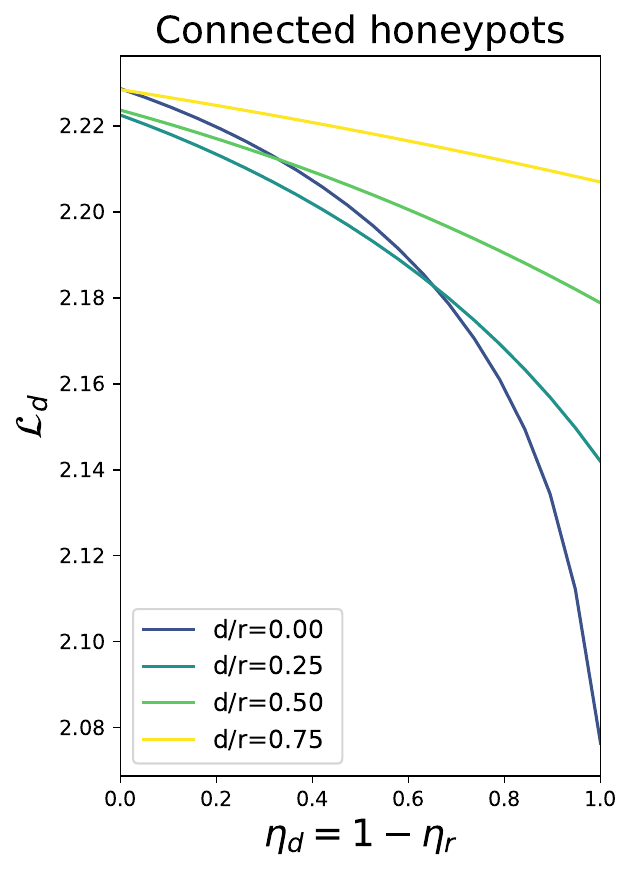}
    \includegraphics[width=0.3\textwidth]{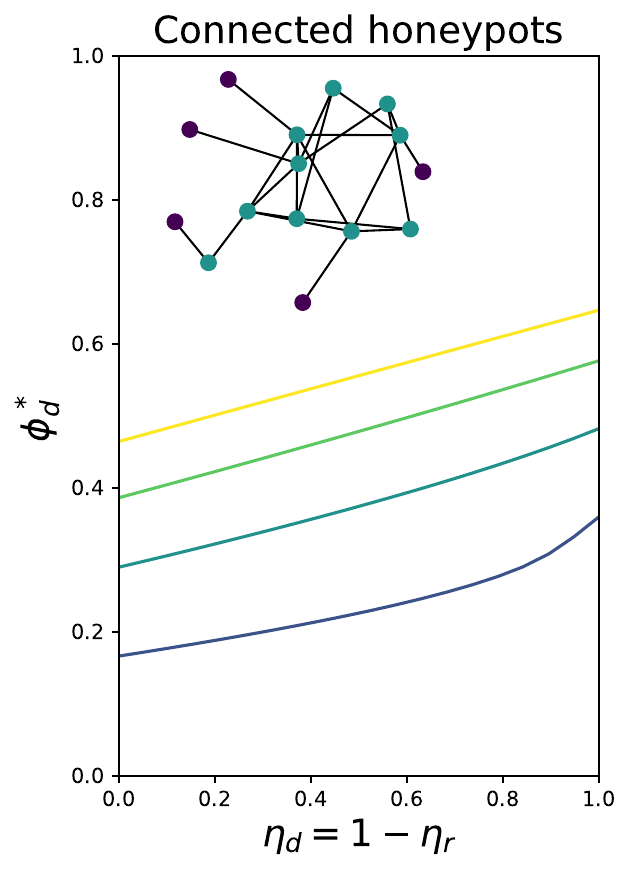}
    \includegraphics[width=0.3\textwidth]{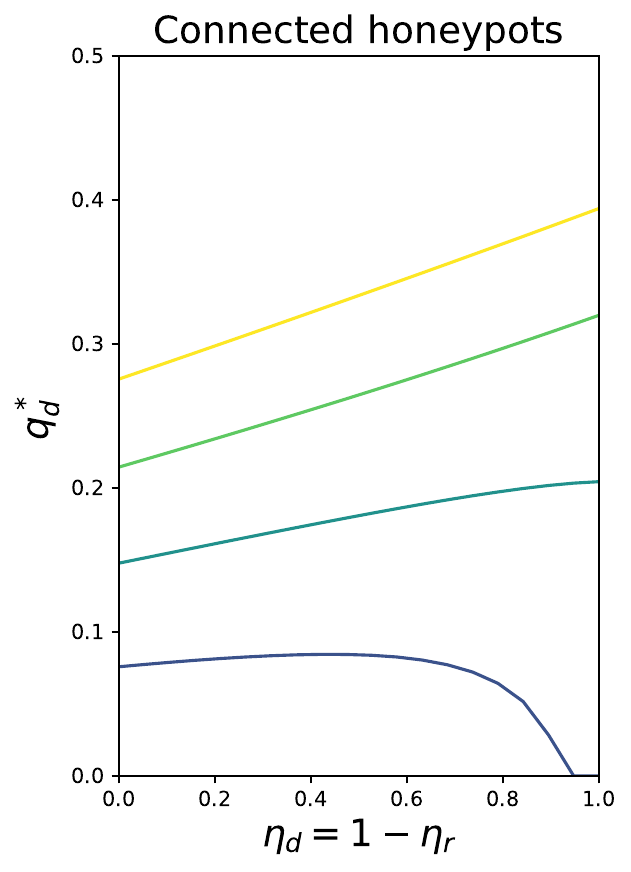}
    \caption{Cyber deception in a network with $r$ nodes and $d$ honeypots with attacker perceived value $\eta_d=1-\eta_r\in[0,1]$ connected to the system. The effectiveness of the strategy $\mathcal{L}_d$ (left panel), the   deception level $\phi^*_d$ (central panel) and the cost of deception $q^*_d=\sum_{i\in N_d} q_i^*$ are shown for increasing fraction of honeypots.}.\label{fig:connected_honeypoint}
    \end{figure}

\begin{figure}[h]
    \centering
    \includegraphics[width=0.49\textwidth]{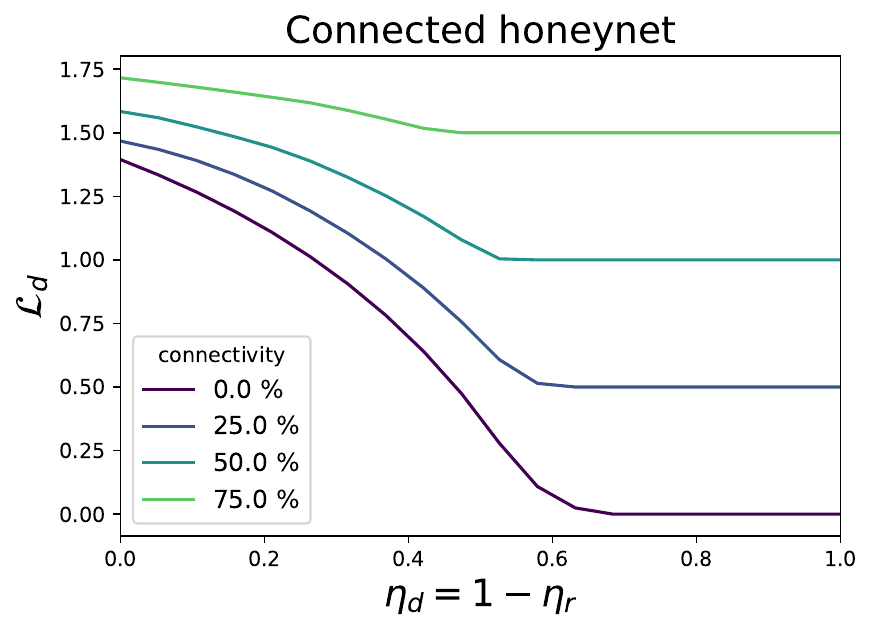}
    \includegraphics[width=0.49\textwidth]{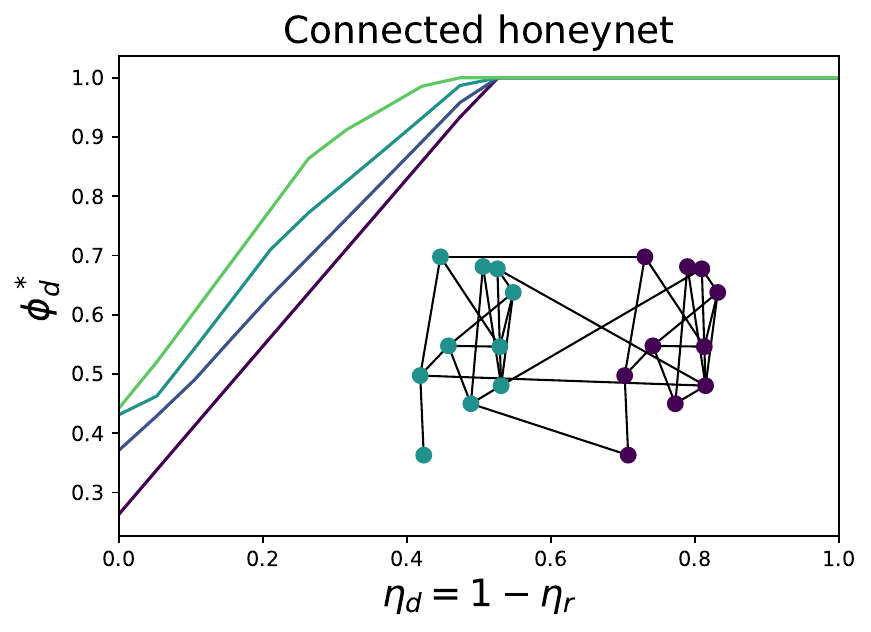}
    \caption{Cyber deception in a network with $r$ nodes and a honeynet that replicates the original system  with attacker perceived value $\eta_d=1-\eta_r\in[0,1]$. The honeynet is connected to the original network. The deception level $\phi^*_d=\sum_{i\in N_d} \phi^*_i$ (right panel) and the effectiveness of the strategy (left panel) are shown for increasing connectivity, i.e. fraction of links connecting nodes in $N_d$ to nodes in $N_r$.}\label{fig:connected_honeynet}
    \end{figure}

To model structural deception techniques, we introduce a partition of the nodes $\mathbb{N}_n=\{1,\ldots, n\}$  into two disjoint subsets $N_r,N_d\subset \mathbb{N}_n$, with $N_r\cap N_d = \emptyset$ and $\mathbb{N}_n=N_r\cup N_d$. Nodes in $N_r$ correspond to genuine components of the system while $N_d$ represents the set of artificially deployed elements. The restriction $\mathcal{G}[N_d]$ of the network to the subset of decoy nodes is called honeynet, while a node $i\in N_d$ that is isolated in $\mathcal{G}[N_d]$ is called honeypot. We assume that the defender assigns value only to real nodes, i.e.
$$ z_i>0 \ \ \implies \ \  i\in N_r,$$
while the attacker may attribute non‑zero value also to decoy nodes. In Fig. (\ref{fig:isolated_honeypoint}-\ref{fig:connected_honeynet}) we considered uniform $\bm{z}=\bm{1}$ and different value profiles $$\bm{\eta}= \eta_r \ \bm{1}_{N_r} + \eta_d\ \bm{1}_{N_d},$$
where $\boldsymbol{1}_{N_r}$ and $\boldsymbol{1}_{N_d}$ denote the indicator vectors of the real‑node and decoy‑node subsets, respectively. The parameter $\eta_d=1-\eta_r \in [0,1]$ represents the perceived value assigned to each node of the honeynet, and therefore quantifies the level of asymmetry in the scenario. We define the \textit{deception level} $\phi^*_d=\sum_{i\in N_d} \phi^*_i $ as the probability at equilibrium that the target seed is in the honeynet, and the \textit{cost of deception} $q^*_d=\sum_{i\in N_d} q_i^*$ as the amount of investment used to defend decoy nodes. Finally we quantify the effectiveness of the deception strategy by measuring how much the defender’s loss decreases relative to the scenario without decoy nodes.

In  Fig.(\ref{fig:isolated_honeypoint}) we consider a system composed of a connected network of $r=|N_r|$ nodes and $d=|N_d|$ isolated honeypots. This allows us to study how the deception effectiveness depends on the proportion $d/r$ of decoy nodes in the system. As expected the deception level $\phi^*_d$ increases as the number of honeypots increases and as their perceived attractiveness $\eta_d$ becomes higher.  Since the honeypots are isolated, an attack against them cannot propagate to the real network and therefore does not contribute to systemic risk. 
As a consequence, the defender is not required to commit resources on $N_d$ (the cost of deception is always zero) and $\mathcal{L}_d$ decreases with $d$ and $\eta_d$.

The situation changes when the honeypots are connected to the real network (see Fig.(\ref{fig:connected_honeypoint})). In this case, the defender must allocate part of the budget to prevent an infection originating in a decoy node from propagating back into the system, namely to construct a \textit{honeywall}. As a consequence, when the number of honeypots increases, the additional cost required to sustain the deception strategy $q^*_d$ outweighs the benefit of the reduced risk, and the defender’s loss function increases.

The previous effect becomes more pronounced as the number of connections between $N_r$ e $N_d$ increases. In  Fig. (\ref{fig:connected_honeynet}), it is shown the case of a honeynet that is an exact replica of the real network, i.e. $r=d$ and $\mathcal{G}[N_r]\equiv \mathcal{G}[N_d]$, where we progressively increase the number of links between the two subsets. It is evident how the effectiveness of deception deteriorates with the increasing connectivity of the honeynet. Interestingly, even when $\eta_{d} = 0$, the presence of an additional deployed subsystem  that is well connected to the original network increases the cyber risk on $N_r$, since the extra nodes can act as a stepping stone through which the infection can propagate back into the real system.

\subsection{Contagion Dynamics}\label{sec:dyn}

The proposed optimal strategies are derived within a game-theoretic framework that bypasses explicit modeling of contagion dynamics in favor of a steady-state approach. Specifically, we assume the epidemic has reached its equilibrium, meaning the infection has fully propagated through the transmission subnetwork defined by the susceptible nodes.

In this static framework, widely utilized in site percolation problems \cite{barrat2008dynamical} and already considered in \cite{aspnes2006inoculation,acemoglu2016network}, the temporal evolution of a contagion is abstracted into a geometric reachability problem. For a single realization of susceptible nodes $\bm{X}$, a node $i$ is infected if, and only if, there exists a path of susceptible nodes connecting it to the initial seed $s$ in the trasmission subnetwork $\mathcal{T}(\bm{X})$. This condition is formally equivalent to the asymptotic state of a Susceptible-Infected (SI) model evolving on the same subnetwork, where at each time step the neighbors of an infected node become infected with a certain rate $\beta$.  Given that the SI model lacks a recovery mechanism, for $t \to \infty$ the infection will, with probability 1, reach every node within the connected component of the seed, regardless of the specific transmission rate $\beta > 0$.

The risk measure $\mathcal{R}_i$ is defined as the average of the infection probability over all possible realizations of node susceptibilities. This quantity can, in turn, be mapped onto the asymptotic infection probability of a heterogeneous SIR (Susceptible-Infected-Removed) model. In this model, each neighbor of an infected node is either successfully infected with probability $(1-q_i)$ or becomes permanently removed (immunized) with probability $q_i$. Consequently, the static clusters of site percolation provide an exact representation of the final outbreak size and the distribution of infected individuals in the long-time limit of the corresponding SIR process.  Within this mapping, as already discussed in Section \ref{sec:path}, the risk measure $\mathcal{R}^L_i$ becomes an estimate of the infection probability of $i$ after exactly $L$ time steps of SIR dynamics.

However, our results can be validated against alternative contagion dynamics, allowing us to assess their robustness to potential model misspecifications. To this end, we consider a benchmark class of epidemic models that describe the temporal evolution of infections across the network. 
Specifically, we consider a Markovian dynamics for the state evolution of nodes. The state at time $t$ is defined by a vector $\bm{\sigma}^t\in\{0,1\}^n$, where $\sigma^t_i$ is equal to one if the $i$-th node of the network is infected at time $t$, and zero otherwise. 

We consider a setting in which, in the absence of security investments, each susceptible node ($(\sigma^t_i=0)$) faces a probability of infection proportional to $\beta$ at each time step, due to contact with each infected neighbors. At the same time, each infected node has a probability $\gamma$ to recover thanks to its baseline security defenses. We therefore consider a general class of possible contagion dynamics defined by the transition rule
\begin{equation}\label{eq:rule}
    \sigma^{t+1}_i = \Theta\left[ \sigma^t_i (1-\gamma-u) +  (1-\sigma^t_i )\left(\frac{\beta}{d_i} \sum_{j=1}^n A_{ij}\sigma^t_j  - Z \right) \right],
\end{equation}
where $d_i = \sum_j A_{ij}$ is the degree of the $i$-th node, i.e. the number of neighbors, and  $u,Z$ are random variables. In particular,  $u\sim U(0,1)$  and $Z$ is a convex combination of an uniform random variable and a deterministic threshold  $\delta \in [0,1]$, i.e. $Z=\varepsilon \delta + (1-\varepsilon) u$, $\tau\in[0,1]$. 

For $\varepsilon=1$, Eq. (\ref{eq:rule}) represents a threshold contagion model \cite{granovetter1978threshold}, in which a susceptible node becomes infected if the fraction of infected neighbors exceeds a given threshold $\delta$. For $\varepsilon=0$, it reduces to a SIS model \cite{pastor2015epidemic} in which the probability of contagion is 
\begin{equation}
\mathbb{P}\left(\sigma_i^{t+1}=1|\sigma_i^t=0,\bm{\sigma}_{\backslash i}^t\right)= 1- \prod_{i=1}^n(1-\beta A_{ij}\sigma^t_j) = \beta \sum_{j=1}^n A_{ij}\sigma^t_j +o(\beta)
\end{equation}
and the contagion rate $\beta$ has been rescaled with the degree to ensure the r.h.s. term is in $[0,1]$. Finally, setting the recovery rate $\gamma$ to zero yields a classical SI model, where infected nodes remain infected indefinitely. 

We assume that both players can influence the contagion dynamics. The defender mitigates the infection rate by investing in cyber-defense, which we model by rescaling the parameters as $\beta \to (1 - q_i)\beta$. The attacker, on the other hand, determines the initial distribution of threats. Accordingly, we define the initial state of the Markov process as
\begin{equation}
\mathbb{P}(\bm{\sigma}^0) = \prod_{i=1}^n (\phi_i(1-q_i))^{\sigma^0_i}(1-\phi_i(1-q_i))^{1-\sigma^0_i}.
\end{equation}
Here, $\phi_i(1 - q_i)$ represents the probability that node $i$ is initially infected, modulated by the defender's investment level $q_i$.
\begin{figure}[h]
    \centering
    \includegraphics[width=\linewidth]{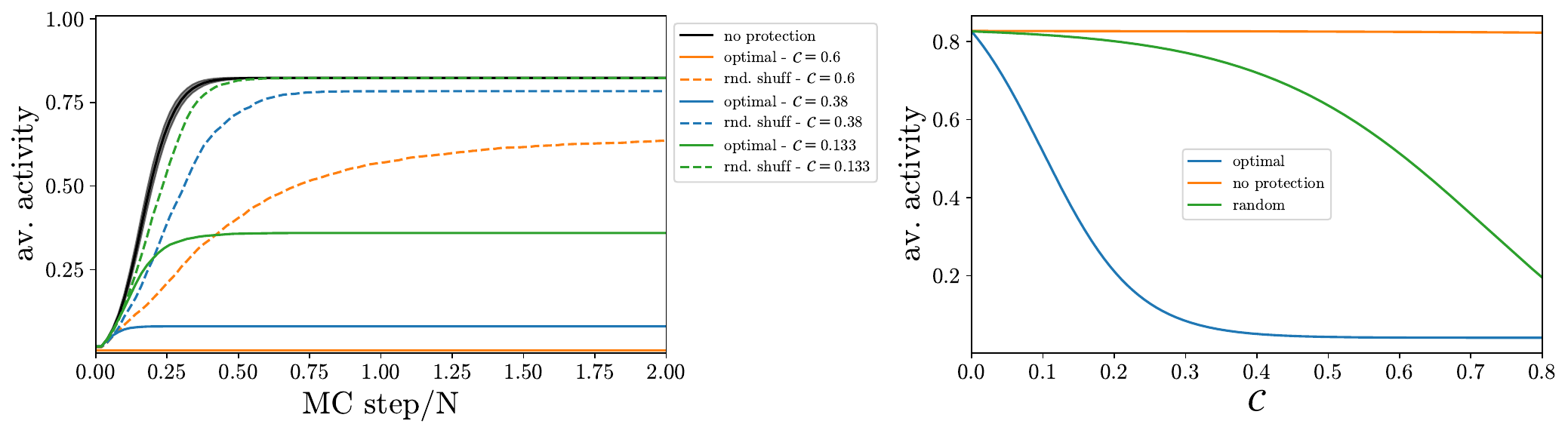} \caption{ SI contagion under different mitigation strategies: no protection, the proposed optimal strategy $\bm{q}^*$ with uniform defender risk profile and a random reshuffling of the same budget. Left Panel: time evolution of the infection's size (average activity). Right Panel: asymptotic infection size as a function of the defender's budget.
    }
    \label{fig:dynamic_uniform}
\end{figure}

\begin{figure}[h]
    \centering
    \includegraphics[width=0.45\linewidth]{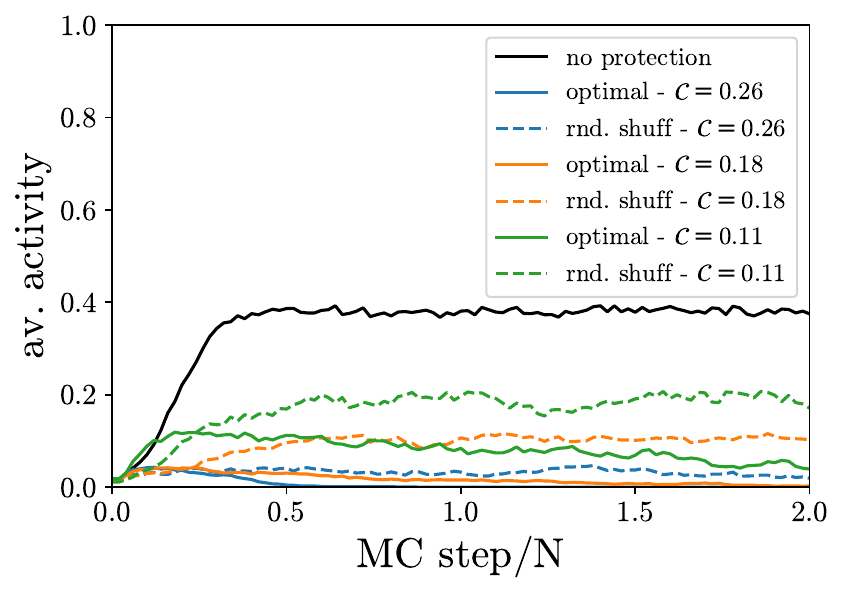}
    \includegraphics[width=0.45\linewidth]{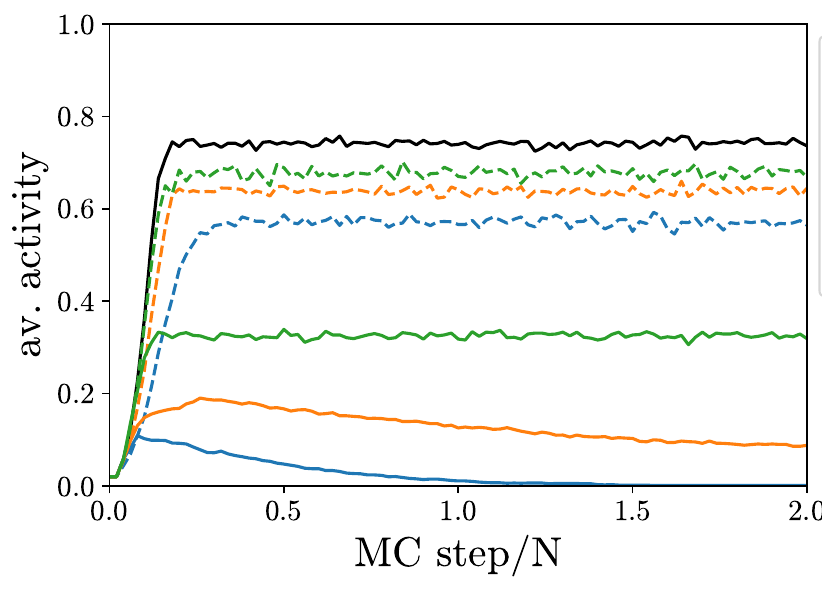}
    \caption{ Infection dynamics under different propagation rules, protection strategies and defender budgets. Left panel: SIS dynamics with $\gamma=0.2$; Right Panel: threshold model with $\gamma=0.2$ and $\delta=0.2$. We compare three  strategies: no protection (black), the proposed optimal strategy $\bm{q}^*$ with uniform defender risk profile (full line) and a random reshuffling of the same budget (dashed line). 
    }
    \label{fig:other_dynamics}
\end{figure}

In Fig. \ref{fig:dynamic_uniform}, we illustrate the effect of a mitigation strategy in random network where the infection evolves according to a SI dynamics on the entire network. We compare the evolution of the fraction of infected nodes in the absence of security investments with that mitigated by the optimal allocation $\bm{q}^*$, derived from the security game equilibrium under uniform value profiles and a fixed budget. The results clearly show that the infection spreads more slowly and remains less extensive. The mitigating effect of this strategy outperforms that of a random reshuffling of the same investment budget, demonstrating its robustness even under contagion mechanism misspecification. The benefit of adopting an optimized strategy is particularly evident in the medium-budget regime, where investments have the potential to reduce infection but are insufficient to fully immunize the system. In this setting, an effective allocation becomes decisive. Fig. \ref{fig:other_dynamics} shows that the previous results remain robust also under contagion dynamics governed by SIS and threshold models.

\begin{figure}[h]
    \centering
    \includegraphics[width=0.45\linewidth]{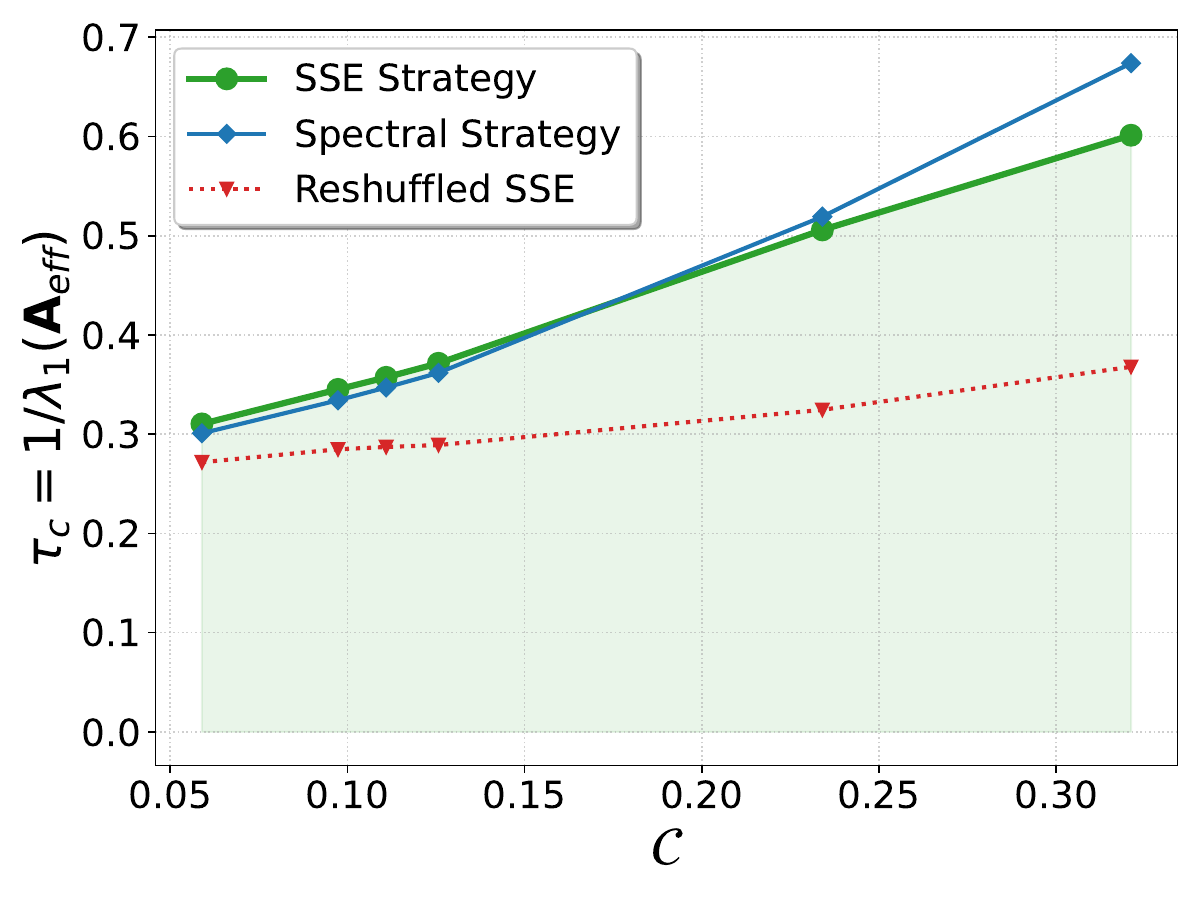}
    \includegraphics[width=0.45\linewidth]{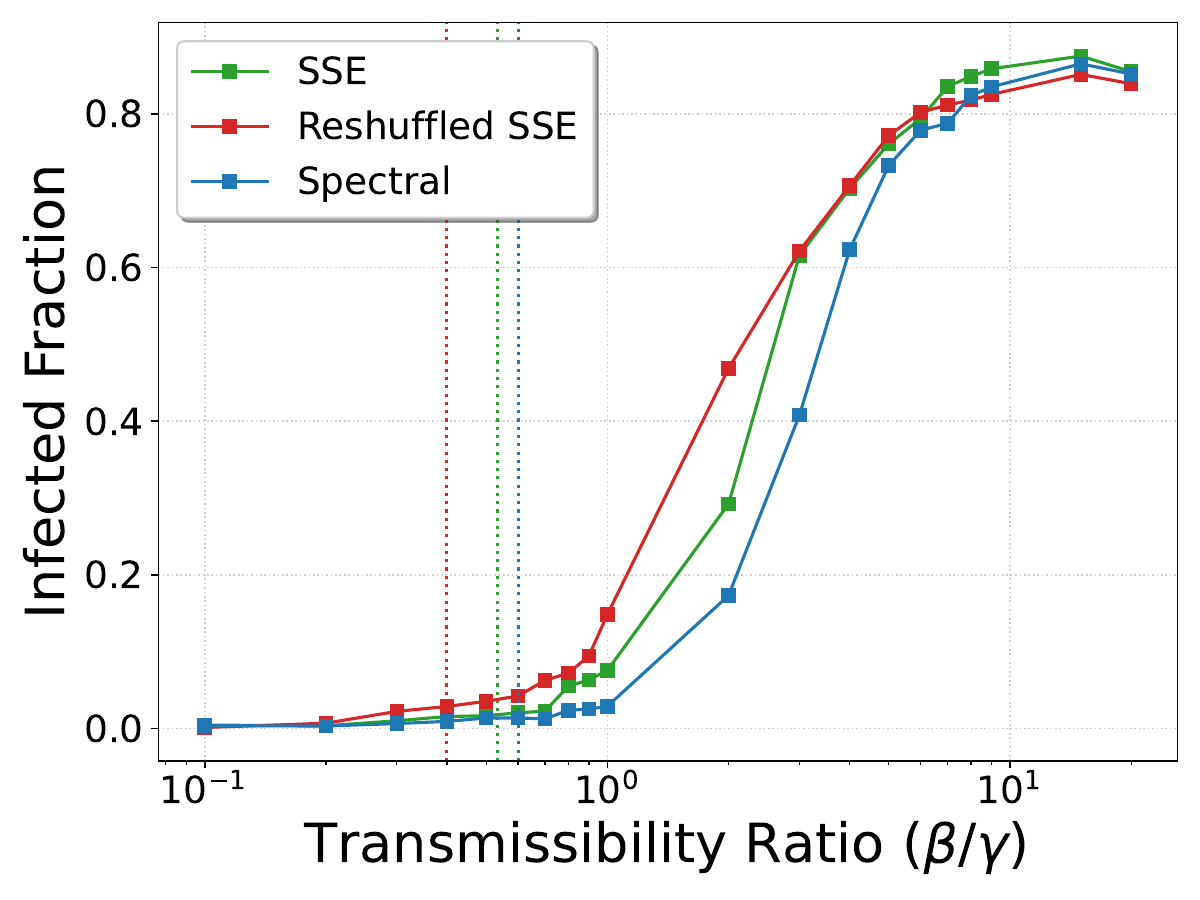}
    \caption{ Optimal defending strategy in an Erdos-Renyi random graph under SIS contagion dynamics. Left Panel: critical transmissibility ratio $\tau_c=1/\lambda_1(\bm{A}_{eff})$, $(\bm{A}_{eff})_{ij}=(1-q_i)A_{ij}$,  for the SSE strategy, a random reshuffle and the optimal spectral strategy, shown as a function of the available budget $\mathcal{C}$. Right Panel: the role of $\tau_c$ (vertical lines) in shaping the phase transition governing the long-term extent of contagion.   
    }
    \label{fig:dinamic_crit}
\end{figure}

Most epidemiological models are characterized by a well-defined threshold parameter, the basic reproduction number $R_0$, which dictates the long-term behavior of the contagion: if $R_0 > 1$, the epidemic becomes systemic, whereas if $R_0 < 1$, it is destined to vanish (see right panel of Fig.\ref{fig:dinamic_crit}). In a networked population, this quantity is typically expressed as the product of the intrinsic disease parameters and a structural term that accounts for the network topology. For instance, in the SIS model, the reproduction number is given by $R_0 = \tau\cdot \lambda_1(\bm{A})$, where $\tau = \beta/\gamma$ represents the effective spreading rate, defined by the ratio between the infection rate $\beta$ and the recovery rate $\gamma$, and $\lambda_1(\bm{A})$ is the spectral radius of the adjacency matrix. This defines the epidemic threshold $\tau_c = 1/\lambda_1(\bm{A})$, the critical infectivity value above which the pathogen can reach an endemic state (in SIS) or trigger a large-scale outbreak (in SIR). Topologies with a smaller $\lambda_1(\bm{A})$ are inherently less vulnerable to contagion. Consequently, a standard optimal defense strategy consists of allocating resources $\mathbf{q}$ to minimize the spectral radius of the effective adjacency matrix $(1-q_i)A_{ij}$.

In Fig. \ref{fig:dinamic_crit}, we compare the epidemic thresholds  $\tau_c(\bm{q})$ obtained using our proposed strategy against the optimal one.  As illustrated in the figure, it is evident that the proposed strategy, despite not being explicitly designed to maximize the epidemic threshold, provides a remarkably close approximation to the optimal spectral defense, particularly in the low-budget regime. Conversely, a reshuffled strategy, which redistributes the same defensive resources randomly across the network, yields significantly poorer results, highlighting the importance of the topological information captured by our approach.

It is important to highlight a fundamental distinction: while spectral minimization primarily focuses on shifting the epidemic threshold $\tau_c$ to prevent the onset of an outbreak, our reachability-based framework is designed to limit the expected magnitude of the contagion once the system is in the supercritical regime. In other words, the spectral method maximizes the global stability of the healthy state, whereas our approach offers a more direct assessment of the expected final outbreak size.

Finally, it is crucial to emphasize the versatility of our framework regarding heterogeneous valuations of the nodes. While spectral approaches are primarily designed to inhibit the onset of an epidemic by treating the network as a uniform structural challenge, they lack a direct counterpart for scenarios involving diverse node valuations, information asymmetry and strategic behaviors. A key advantage of our proposal is that, in addition to yielding results comparable to dynamic epidemiological spectral methods in homogeneous settings, it can be seamlessly generalized to more realistic contexts where heterogeneity is a dominant factor.

\section*{Conclusions and Policy Implications} \label{sec:conclusions}

In this paper, we considered a general class of security games with contagion to study the optimal allocation of cybersecurity investments across the nodes of a networked system subjected to strategic attacks. Our analysis focused on the role of asymmetry between the risk profiles targeted by the attacker and the defender, as well as their heterogeneity.

We showed that, in a low-budget regime, the Stackelberg equilibria of the game can be approximated using simple network metrics that capture each node’s ability to protect its neighbors and its value relative to the players’ objectives. A key contribution of the analysis is that these protection metrics emerge endogenously from the equilibrium characterization rather than being imposed ex ante. We also proposed a risk measure based on the number of contagion paths, making the optimization problem scalable for larger networks.

From numerical experiments, we showed that, unlike standard security games, the presence of contagion mechanisms fosters the emergence of optimal strategies involving cyber deception. We analyzed efficient frontiers to compare systemic risk across networks with different topologies and to highlight how asymmetry and heterogeneity in risk profiles shape systemic resilience. Finally, we assessed the robustness of the proposed optimal strategies under alternative contagion dynamics.

Beyond the theoretical characterization of equilibrium strategies, the results have direct implications for cybersecurity risk management. A central finding of the paper is that cyber-risk heterogeneity significantly affects the efficient allocation of defensive resources. In particular, optimal cybersecurity investments are determined not only by the position of nodes within the network topology, but also by their economic relevance and by the objectives pursued by potential attackers. As a consequence, relying exclusively on standard network indicators such as connectivity or centrality may lead to inefficient allocations of cybersecurity resources. More generally, our findings show that neglecting heterogeneity may generate systematic misallocation of cybersecurity investments, reducing overall system resilience.

From a managerial perspective, effective cyber-risk management should therefore integrate network analysis with asset valuation, business-impact assessments, and operational risk measures. Organizations managing complex digital infrastructures should account for both the systemic role of individual assets and their contribution to organizational objectives when defining investment priorities. Ignoring these dimensions may result in overprotection of highly connected assets and underprotection of economically critical components.

From a policy perspective, our findings highlight the limitations of uniform cybersecurity requirements applied equally across all assets or organizational units. In heterogeneous systems, resilience can be strengthened through risk-based approaches that explicitly account for differences in criticality and systemic importance. This consideration is particularly relevant for financial institutions, payment infrastructures, cloud-service providers, and other critical digital ecosystems where local cyber incidents may generate broader systemic consequences.

More broadly, the analysis emphasizes the role of information asymmetries in cybersecurity. Differences between the value assigned to assets by attackers and defenders substantially affect equilibrium outcomes and resource allocation decisions. In this context, cyber-deception mechanisms can be interpreted as tools for strategically influencing attackers’ beliefs and redirecting attacks away from genuinely critical assets. This interpretation links cybersecurity investments to broader economic themes related to asymmetric information, signaling, and strategic behavior under uncertainty.

Future perspectives will consider scenarios with incomplete or asymmetric information, examining how equilibrium strategies shift when defenders lack knowledge of the attacker’s objectives and vice versa. Moreover, both attacker and defender are assumed to have perfect information about the system’s architecture, which simplifies analysis but reduces realism. Assuming partial information on network connections, especially for the attacker, would represent an important extension of the current framework.

Another promising extension is a dynamic game formulation in which defenders adapt their strategies as contagion unfolds. This could involve reallocating defenses across nodes or modifying the network topology, for example by disabling critical links. Coupling contagion dynamics with an iterative two-player game operating on different timescales would highlight a key missing element: the role of timing and the risks associated with delayed responses. Finally, future research could investigate the design of optimal information and deception policies, in which defenders strategically shape attackers’ beliefs while balancing implementation costs and operational constraints.

\section{Appendix A: Proof of Theorem \ref{thm:approxSSE}}\label{sec:app}

Before proving Theorem \ref{thm:approxSSE}, let us state some technical results. Throughout this appendix, let us denote $\mathbb P(\q, \bm\phi;\bm A)$ simply by $\mathbb P(\q, \bm\phi)$. By the definition \label{eq:P_i(q,phi;A)}\!\!, we have the following decomposition of the probability that the node $i$ becomes infected:
\begin{equation}
    \mathbb P_i(\q,\bm\phi)=(1-q_i)\tilde{\mathbb P}_i(\q_{-i},\bm\phi),
\end{equation}
where $(1-q_i)$ is the probability that the node $i$ is susceptible and the marginal $\tilde{\mathbb P}_i(\q_{-i},\bm\phi)$ is the probability that the infection reaches the node $i$ then it does not depends on the pointwise security investment $q_i$, so $\q_{-i}$ denotes the security profile $\q$ without the entry $q_i$. Therefore, the probability that the node $j$, with $j\not =i$, becomes infected can be written as (Proposition 2, \cite{acemoglu2016network}):
\begin{equation}
    \tilde {\mathbb P}_j(\q_{-j},\bm\phi)=\tilde{\mathbb P}_j(\q_{-\{i,j\}},\bm \phi)+(1-q_i)Q_{ji}(\q_{-\{i,j\}},\bm\phi),
\end{equation}
where $\tilde{\mathbb P}(\q_{-\{i,j\}},\bm \phi)$ is the probability that the infection reaches the node $j$ through a path that does not contain the node $i$ and $Q_{ji}(\q_{-\{i,j\}},\bm\phi)$ is the probability that the infection reaches the node $j$ through a path that contains the node $i$ conditioned $i$ being susceptible. As a consequence, we obtain the following:
\begin{equation}\label{eq:partial_iP_j}
       \frac{\partial}{\partial q_i}\mathbb P_j(\q,\bm\phi)=-\delta_{ij}\tilde{\mathbb P}_i(\q,\bm\phi) - (1-\delta_{ij})(1-q_j)Q_{ji}(\q,\bm\phi),
\end{equation}
where we have omitted the subscripts of $\q_{-i}$ and $\q_{-\{i,j\}}$ for convenience of notation.

\begin{lemma}\label{lemma1}
Given the security profile $\q$ and the attack distribution $\bm \phi$, we have:
\begin{equation}\label{eq:approx P}
      \tilde{\mathbb P}_i(\q,\boldsymbol{\phi})=1-\sum_{j\neq i}\sum_t \phi_ta_{it}^j q_j + o(\|\q\|_\infty), 
      \end{equation}
      \begin{equation}\label{eq:approx Q}
 (1-q_j)Q_{ji}(\q,\bm \phi)=\sum_t \phi_t \left( a^i_{jt}+\sum_{k\neq i}(b_{jt}^{(i,k)}-a^i_{jt}a^k_{jt})q_k\right)+o(\|\q\|_{\infty}).
\end{equation}

\end{lemma}
\begin{proof}
The claim is a direct generalization of Proposition 4 in \cite{acemoglu2016network}.
\end{proof}

\begin{proof}[Theorem \ref{thm:approxSSE}]

Let us consider the optimization problem over the attacker's utility, with risk function defined in \eqref{eq:risk1} and quadratic cost function $\mathcal C_a(\bm\phi)=\frac{1}{2}\sum_{i=1}^n \phi_i^2$, in order to obtain the best response to the defender's strategy $\bm q$:
    \begin{equation}
        \max_{\bm \phi} \mathcal U_a(\bm\phi;\q)=\max_{\bm \phi}\left(\sum_{i=1}^n\eta_i\mathbb P_i(\q,\bm\phi)-\frac{\theta}{2}\sum_{i=1}^n\phi_i^2 \right)
    \end{equation}
    subjected to the constraints $\sum_i \phi_i=1$ and $\phi_i\geq 0$ $\forall i=1,\ldots, n$.
    Let $L=L(\bm q, \bm\phi, \lambda,\mu)$ be the Lagrangian function, with parameters $\lambda \in \mathbb{R}$ and $\mu_i \in \mathbb{R}^+$ for all $i=1,\ldots,n$, related to the considered constrained optimization problem: 
    \begin{equation*}
         L(\q, \bm\phi, \lambda,\mu)= \sum_{i=1}^{n} \eta_i  \mathbb P_i(\bm q,\bm\phi) - \frac{\theta}{2}\sum_{i=1}^{n}\phi_i^2 +\lambda\left(\sum_{i=1}^{n} \phi_i -1 \right) +\sum_{i=1}^{n} \mu_i \phi_i.
        \end{equation*}
        Requiring the optimality condition:
         $$\frac{\partial}{\partial \phi_i}L(\q,\bm\phi, \lambda,\mu)=\mathbb P_i(\q,\bm \eta)-\theta \phi_i+\lambda+\mu_i=0, \quad \forall i=1,...,n,$$ 
          where 
 \begin{equation}\label{eq:P(eta)}
     \mathbb P_i(\q,\bm \eta)=\sum_{s=1}^n\eta_s\mathbb E_{\bm X}[\mathbb I(i\sim s\in \mathcal T(\bm X))],
 \end{equation}
        and imposing the constraints, we have that the best response $\bm \phi^*(\q)=(\phi_1^*(\q),...,\phi^*_n(\q))$ of the attacker is given by: 
\begin{equation}
\label{phi_star}
  \phi_i^*(\q)=\frac{1}{n}+\frac{1}{\theta}\left(\mathbb P_i(\q,\bm \eta)-\frac{1}{n}\sum_{j=1}^n \mathbb P_j(\q,\bm \eta)\right),\quad i=1,...,n.
 \end{equation}
Let us remark that the function in \eqref{eq:P(eta)} directly relies the definition of the probability \eqref{eq:P_i}, however $\mathbb P_i(\q,\bm \eta)$ is not a probability value since the weights $\eta_1,\dots,\eta_n$ are not convex coefficients in general.

In order to estimate the optimal strategy $\q^*$ of the defender, we are now interested in minimizing the defender's loss, with quadratic cost function $\mathcal C_d=\frac{1}{2}\sum_{i=1}^n q_i^2$, for $\bm\phi=\bm \phi^*$: 
 \begin{align}\label{eq:optimalL_d}
     0&=\frac{\partial}{\partial q_i} \mathcal L_d(\q;\bm \phi^*)=\frac{\partial}{\partial q_i}\left(\sum_{s=1}^nz_s\mathcal R_s(\q,\bm \phi^*;\bm A)+\frac{\alpha}{2}\sum_{s=1}^nq_s^2 \right).\end{align}

\noindent
Let $\bm e_s$ be the $n$-dimensional vector with 1 in the $s$-th entry and 0 elsewhere and $\bm 1_n$ is the $n$-dimensional vector with $1/n$ in every entry. By definition \eqref{eq:P_i}, we have:
$$\mathbb P_i(\q,\bm e_k)=\mathbb E_{\bm X}[\mathbb I(i\sim k\in \mathcal T(\bm X))]. 
$$
Therefore, the derivative of $\mathcal L_d$ is
\begin{align}\label{eq:partialL_d}
\frac{\partial \mathcal{L}_d}{\partial q_i}&= \alpha q_i + \sum_{k,s} z_s \phi^*_k \frac{\partial \mathbb P_s(\q,\bm{e}_k)}{\partial q_i}+\sum_{k,s} z_s \mathbb P_s(\q,\bm{e}_k) \frac{\partial \phi_k^*(\q)}{\partial q_i} \nonumber \\
&:= \alpha q_i +x_i + y_i.
\end{align}
The term $x_i$ is exactly the derivative computed for the social equilibrium in the random case, evaluated at $\bm{\phi}^*$, while the term $y_i$  is due to the strategic nature of the problem and involves the derivative of $\bm{\phi}^*=\bm \phi^*(\q)$. Let us compute the two terms applying \eqref{eq:partial_iP_j}:
\begin{align}
x_i&= \sum_{k,s} z_s \phi^*_k \frac{\partial \mathbb P_s(\q,\bm{e}_k)}{\partial q_i}\nonumber\\
&= -z_i \tilde{\mathbb P}_i(\q,\bm{\phi}^*)-\sum_{s\neq i} z_s (1-q_s) Q_{si}(\q,\bm{\phi}^*)\nonumber\\
&= - z_i \left( 1- \sum_{j\neq i} a^j_i(\bm{\phi}^*) q_j\right)- \\
&-\left( \sum_{j\neq i} z_j a^i_j(\bm{\phi}^*) +\sum_{j\neq i} z_j \sum_s\sum_{k\neq i} \phi^*_s \left(b_{js}^{(i,k)}-a^i_{js}a^k_{js}\right) q_k \right) +o(\|\q\|_\infty)\nonumber \\
&= -\sum_j z_j a^i_j(\bm{\phi}^*) - \sum_{j\neq i} \sum_{sk} \phi^*_sz_k \left(b^{(ij)}_{ks}-a^i_{ks}a^j_{ks}\right) q_j+o(\|\q\|_\infty)\nonumber\\
&= -a^i(\bm{\phi}^*,\bm z) - \sum_{j\neq i} b_{ij}(\bm{\phi}^*,\bm z) q_j +o(\|\q\|_\infty)
\end{align}
where we used the approximations for $\tilde{\mathbb P}$ and $Q$ provided by \autoref{lemma1}, $a^j_i=a^j_i(\bm\phi)$, $a^i=a^i(\bm\phi,\bm z)$, and $b_{ij}=b_{ij}(\bm{\phi},\bm z)$ are defined in \eqref{eq:a^j_i}, \eqref{eq:a^i}, and \eqref{eq:b_ij} respectively.
By inserting $\bm{\phi}^*(q)$, given in \eqref{phi_star}, and observing that 
$$\phi^*_k(\q)=1/n +\nabla\phi^*_k(0)\cdot \q +o(\q),$$ 
we have:
\begin{equation}\label{eq:A}
x_i= -a^i(1/n,\bm z)-\sum_{j\neq i} b_{ij}(1/n,\bm z) q_j - \sum_{jk} a^i_{jk} z_j \sum_s \frac{\partial \phi^*_k}{\partial {q_s}}\big |_{\q=0} q_s +o(\|\q\|_\infty),
\end{equation}
where the first two terms are exactly the ones appearing in the random case.
The term $y_i$ is 
\begin{align}\label{eq:y}
y_i&=\sum_{k,s} z_s \mathbb P_s(\q,\bm{e}_k) \frac{\partial \phi^*_k(\q)}{\partial {q_i}} =\sum_{k,s} z_s (1-q_s) \tilde{\mathbb P}_s(\q,\bm{e}_k) \frac{\partial \phi^*_k(\q)}{\partial {q_i}}\nonumber \\
&= \sum_{ks} z_s (1-q_s)\left(1-\sum_{j\neq s} a^j_{sk} q_j \right) \frac{\partial \phi^*_k(\q)}{\partial {q_i}} +o(\|\q\|_\infty)\nonumber \\
&= \sum_{ks} z_s (1-q_s) \frac{\partial \phi^*_k(\q)}{\partial {q_i}}- \sum_{ks} z_s \sum_{j\neq s} a^j_{sk} q_j  \frac{\partial \phi^*_k(\q)}{\partial {q_i}} +o(\|\q\|_\infty) \nonumber \\
&= - \sum_{ks} z_s \sum_{j\neq s} a^j_{sk} q_j  \frac{\partial \phi^*_k}{\partial {q_i}}\big |_{\q=0}  +o(\|\q\|_\infty), 
\end{align}
where we used that $\sum_k \phi^*_k=1$ and that only  $ \partial_{q_i} \phi^*_k(0)$ contributes to the leading order. To make explicit $x_i$ and $y_i$, we just need to compute the term
\begin{align}
 \frac{\partial\phi_k^*}{\partial q_i}\big|_{\q=0}&= \frac{\partial}{\partial q_i} \left( \frac 1 n +\frac 1 \theta \left( \mathbb P_k(\q,\bm{\eta})-\frac 1 n \sum_j \mathbb P_j(\q,\bm{\eta}) \right)    \right)\big|_{\q=0}\nonumber \\
 &=  \frac 1 \theta \left( \frac{\partial \mathbb P_k(\q,\bm{\eta})}{\partial q_i} -\frac 1 n \sum_j \frac{\partial \mathbb P_j(\q,\bm{\eta})}{\partial q_i} \right)\big|_{\q=0}\nonumber\\
 &=  \frac 1 \theta \left( -\delta_{ik}\tilde{\mathbb P}_i(\q,\bm{\eta})-(1-\delta_{ik}) (1-q_k) Q_{ki}(\q,\bm{\eta}) \right. \nonumber \\
 &\left .  -\frac 1 n \sum_j \left(
 -\delta_{ij}\tilde{\mathbb P}_i(\q,\bm{\eta})-(1-\delta_{ij}) (1-q_j) Q_{ji}(\q,\bm{\eta})    \right)     \right)\big|_{\q=0}\nonumber \\
 &=\frac 1 \theta \left( -\delta_{ki} -(1-\delta _{ki}) a^i_k(\bm{\eta}) +\frac 1 n \sum_j \left( \delta_{ji} +(1-\delta _{ji}) a^i_j(\bm{\eta})   \right)  \right)\nonumber \\
&= \frac 1 \theta \left( a^i(1/n,\bm{\eta}) -a^i_k(\bm{\eta})\right).
\end{align}
By inserting it into \eqref{eq:y}, we have:
\begin{align}
y_i&=  \frac 1 \theta \sum_{ks} z_s \sum_{j\neq s} a^j_{sk} q_j  \left( a^i_k(\bm{\eta}) - a^i(1/n,\bm{\eta})\right)   +o(\|\q\|_\infty)\nonumber \\
&= \frac 1 \theta \sum_{k,s,j} z_s a^j_{sk} q_j  \left( a^i_k(\bm{\eta}) - a^i(1/n,\bm{\eta})\right)   +o(\|\q\|_\infty) \nonumber \\
&= \frac 1 \theta \sum_j q_j  \sum_{k} a^j_k(\bm{z})\left( a^i_k(\bm{\eta}) -a^i(1/n,\bm{\eta})    \right ) +o(\|\q\|_\infty)\nonumber \\
&= \frac 1 \theta \sum_j   \left( \sum_{k} a^i_k(\bm{\eta}) a^j_k(\bm{z}) -a^i(1/n,\bm{\eta}) a^j(1/n,\bm{z})   \right)q_j +o(\|\q\|_\infty),\nonumber 
\end{align}
where, in the second line we used that we have zero if $j=s$. 
Using the same type of computaion  in Eq. (\ref{eq:A}) we obtain
\begin{align}
    x_i=& -a^i(1/n,\bm z)-\sum_{j\neq i} b_{ij}(1/n,\bm z) q_j \\
    &+ \frac 1 \theta\sum_j    \left( \sum_{k} a^i_k(\bm{z}) a^j_k(\bm{\eta}) -a^i(1/n,\bm{z}) a^j(1/n,\bm{\eta}) \right) q_j +o(\|\q\|_\infty).
    \end{align}
Putting all together into \eqref{eq:partialL_d} we get:
\begin{align}
\frac{\partial \mathcal{L}_d}{\partial q_i}=&\ \alpha q_i-a^i(1/n,\bm z)\\
&-\sum_{j\neq i} b_{ij}(1/n,\bm z) q_j + \frac 1 \theta\sum_j    \left( \sum_{k} a^i_k(\bm{\eta}) a^j_k(\bm{z}) -a^i(1/n,\bm{\eta}) a^j(1/n,\bm{z}) \right) q_j \nonumber\\
& + \frac 1 \theta \sum_j   \left( \sum_{k} a^i_k(\bm{z}) a^j_k(\bm{\eta}) -a^i(1/n,\bm{z}) a^j(1/n,\bm{\eta})   \right)q_j +o(\|\q\|_\infty)\\\nonumber
=&\ \alpha q_i - s^i(\bm{z}, \bm{\eta}; \bm{p}^1, \bm{p}^2) - \sum_j M_{ij}(\bm{z}, \bm{\eta}; \bm{p}^1, \bm{p}^2) q_j+ o(\|\q\|_\infty),
\end{align}
where the vector $\bm s$ and the matrix $\bm M$ are defined in \eqref{eq:a^ib_ij}, as a consequence the equilibrium $\q^*$ satisfies \eqref{eq:qstar}.

\end{proof}


\bibliographystyle{splncs04}
\bibliography{miabiblio.bib}

\end{document}